\documentstyle[preprint,tighten,aps,prb,epsf,eqsecnum]{revtex}

\begin{document}

\title{Parametric Conductance Correlation for Irregularly Shaped
       Quantum Dots}

\author{Henrik Bruus}

\address{CNRS-CRTBT, 25 Avenue des Martyrs, BP166, F-38042
	 Grenoble C\'edex 9, France}

\author{Caio H. Lewenkopf}

\address{Instituto de F\'{\i}sica, Universidade de S\~ao Paulo, 
           Caixa Postal 66318, 05389-970, S\~ao Paulo, Brazil}

\author{Eduardo R. Mucciolo}

\address{NORDITA, Blegdamsvej 17, DK-2100 Copenhagen {\O}, Denmark}

\date{October 20, 1995 [Phys.\ Rev.\ B, {\bf 53}, 9686 (1996)]}

\maketitle

\begin{abstract}
We propose the autocorrelator of conductance peak heights as a
signature of the underlying chaotic dynamics in quantum dots in the
Coulomb blockade regime. This correlation function is directly
accessible to experiments and its decay width contains interesting
information about the underlying electron dynamics. Analytical results
are derived in the framework of random matrix theory in the regime of
broken time-reversal symmetry. The final expression, upon rescaling,
becomes independent of the details of the system. For the situation
when the external parameter is a variable magnetic field, the
system-dependent, nonuniversal field scaling factor is obtained by a
semiclassical approach. The validity of our findings is confirmed by a
comparison with results of an exact numerical diagonalization of the
conformal billiard threaded by a magnetic flux line.
\end{abstract}

\draft

\narrowtext

\section{Introduction}

During the last five years the fabrication of ballistic
two-dimensional electron gas microstructures in
GaAs/Ga$_{1-x}$Al$_x$As heterostructures has opened up a new field:
the study of quantum manifestations of classical chaos in condensed
matter systems.\cite{ChaosReview} The first theoretical
discussion\cite{Jalabert90} of the so-called quantum chaos in
microstructures dealt with open systems, i.e., cavities on the
micrometer scale or smaller connected to the external world by leads
that admit a continuous flow of electrons. Several successful
experiments followed.\cite{Marcus92,Kel94,Chang94,MJBerry94,Bird94}
Nowadays the theoretical study of quantum chaos is also well developed
for closed microstructures, known as quantum dots.\cite{Kastner92}
These are cavities coupled to the external leads by tunnel barriers
that prevent a continuous flow of electrons and make the electric
charge inside the dot quantized. The striking characteristic of
quantum dots is the appearance of very sharp peaks in the conductance
as a function of the gate voltage. These peaks are roughly equidistant
due to the strong charging effect, but their heights vary randomly by
an order of magnitude or more. The general interest in this
phenomenon, combined with the existence of interesting experimental
data on quantum chaos for closed systems,\cite{Sivan94} motivated
several theoretical
studies.\cite{Jalabert92,Prigodin93,Bruus94,Mucciolo95a,Alhassid95}
These works proposed universal conductance peak distributions for
quantum dots in the Coulomb blockade regime as fingerprints of the
underlying chaotic dynamics. Similar predictions for open
systems\cite{Prigodin93,Baranger94} have been difficult to observe
experimentally\cite{Chang95} because of the strong influence of
dephasing.\cite{Brouwer95}

The initial experiments on Coulomb blockade peak heights in quantum
dots\cite{Kastner92} have paved the way for the first convincing
measurements of statistical distributions of peak
heights.\cite{Chang95,Klein95,Folk95} However, the full distribution
of the conductance peaks is not a trivial quantity to measure. Since
each peak sequence is not very long, good statistics demands great
experimental effort. One possible solution is a true ensemble
averaging, which requires many different samples. In
Ref.~\onlinecite{Bruus94} it was suggested that one could obtain such
an average from a single device by changing its shape {\it in situ}
using external electrostatic potentials. This difficult procedure has
now been tested experimentally in open\cite{Chan95} and closed
systems.\cite{Folk95} Another seemingly simpler solution is to
generate more statistics by varying an applied magnetic field (taking
care to measure at $B$-field steps larger than the conductance
autocorrelation width $\delta B_c$).\cite{Chang95,Klein95}

In order to test more thoroughly quantum manifestations of classical
chaos in quantum dots in the tunneling regime we propose a different
quantity to be investigated: the parametric conductance peak
autocorrelation function. Because it is not difficult to cover a
parametric range which extends over several autocorrelation lengths,
correlation functions need a small number of peaks and should be easy
to obtain experimentally. Moreover, correlators of a fixed eigenstate
are also of theoretical interest. Analytical calculations using random
matrix theory beyond a perturbative approach seem to be very
challenging. In addition, the nonuniversal scaling contains
interesting information about the underlying classical dynamics. To
put the parametric dependence in contact with physical quantities and
develop a satisfactory theoretical treatment of nonuniversal scaling
factors, one has to invoke semiclassical arguments. In this sense the
study of the autocorrelator of conductance peaks is an example where
semiclassical and random matrix theories complement each other in the
understanding of mesoscopic phenomena.

The first step towards the description of the main features of chaos
in quantum dots is to consider the problem of electrons moving in a
cavity. We assume that the Coulomb interaction within the cavity can
be taken into account in a self-consistent way, yielding
noninteracting quasiparticles. As a result of extensive studies of
two-dimensional (2D) dynamical systems, one has learned that when the
classical motion in the cavity is hyperbolic, the quantum spectrum and
the wave functions exhibit certain universal features.\cite{Bohigas91}
Throughout this paper we shall consider a statistical approach which
is valid for chaotic systems and also works fairly well for systems
whose phase space is predominantly chaotic.\cite{Obs1}

The paper is organized as follows. In Sec.~\ref{sec:stochastic} we
present our analytical and numerical treatment based on random matrix
theory (RMT). We concentrate on the case of broken time-reversal
symmetry (TR), but also show numerical results for the TR preserved
case (spin-orbit interactions are always neglected). In
Sec.~\ref{sec:dynamical} we study a dynamical model to illustrate our
findings. There we conjecture that the semiclassical periodic orbit
theory can be used to estimate the typical magnetic field scale of any
correlator of a quantum dot in a realistic situation. We conclude in
Sec.~\ref{sec:conclusions} with a discussion on how the presented
results are robust with respect to dephasing effects and on possible
experimental realizations. Several more technical details are left to
the Appendices.

\section{Stochastic Approach}
\label{sec:stochastic}

One of the basic tools to investigate the conductance in mesoscopic
devices is the Landauer-B\"uttiker formula.\cite{Landauer57,Buettiker}
For a two-lead geometry, where leads are denoted by $L$ (left) and $R$
(right), the conductance $G$ is given by
\begin{equation}
G = \frac{2e^2}{h}\sum_{ab} \left| S_{ab}^{LR}(E_F,\phi) \right|^2 \;
,
\label{Conductance}
\end{equation}
where the sum runs over the channels $\{a\}$ on the lead $L$ and the
channels $\{b\}$ on the lead $R$. The scattering matrix $S(E,\phi)$
connects right and left channels and is evaluated at the energy $E$
and magnetic flux $\phi$. The factor of 2 is due to spin degeneracy.

In the case of quantum dots in the tunneling regime this formula
becomes particularly simple, since the $S$ matrix can be written as
\begin{eqnarray}
\label{eq:Smatrix}
S_{ab}(E,\phi) & = & S^0_{ab}(E,\phi) - i \sum_\nu
\frac{\gamma_{a\nu}\, \gamma_{b\nu}} {E - \varepsilon_\nu +
i\Gamma_\nu/2} +\ O(\langle\Gamma\rangle/\Delta) \;,
\end{eqnarray}
where the matrix elements $\gamma_{a\nu}$ give the probability
amplitude of a state in channel $a$ to couple to the resonance $\nu$
in the dot. The total decay width is given by $\Gamma_\nu = \sum_c
|\gamma_{c\nu}|^2$, where the sum is taken over all open channels
$c$. We shall assume that the contribution of direct processes is very
small and neglect the off-resonance term $S^0$. Equation
(\ref{eq:Smatrix}) is a very good approximation to the $S$ matrix when
the average decay width $\langle\Gamma\rangle$ is much smaller than
the mean spacing between resonant states $\Delta$. This is indeed the
case for the isolated resonances observed in quantum dot
experiments. (Hereafter we use $\Delta$ to denote the level spacing
caused solely by single-particle states within the cavity, as opposed
to the spacing between conductance peaks observed in the experiments.)
When the condition $\langle\Gamma\rangle/{\Delta} \ll 1$ is not met,
unitarity corrections to Eq.~(\ref{eq:Smatrix}) become important and
they demand a different parametrization of the $S$ matrix (see, for
example, Ref. \onlinecite{Lewenkopf91}).

The matrix elements in Eq.~(\ref{eq:Smatrix}) can be obtained by
invoking the $R$-matrix formalism.\cite{Lane58} Such an approach was
originally proposed in the study of nuclear resonance scattering and
has more recently been successfully applied in the context of
mesoscopic physics.\cite{Jalabert92,Zirnbauer93} The partial decay
amplitude $\gamma_{c\nu}$ is given by
\begin{equation}
\gamma_{c\nu} = \sqrt{\frac{\hbar^2}{2m}} \int
\!ds\,\chi_c^*(\bbox{r})\psi_\nu(\bbox{r})\;,
\label{eq:gamma}
\end{equation}
where $\psi_\nu$ is the eigenfunction of the resonance $\nu$ {\it
inside} the dot with the appropriate boundary condition and $\chi_c$
is the wave function in the channel $c$ at energy $\varepsilon_\nu$.
The integration is done over the contact region between lead and
cavity. A useful quantity for the forthcoming discussion is the
partial decay width, defined as $\Gamma_{c\nu} =
|\gamma_{c\nu}|^2$. We remark that Eq.~(\ref{eq:gamma}), as it stands,
does not contain barrier penetration factors. Throughout this paper we
will assume that penetration factors are channel independent and
therefore will influence only the average decay width
$\langle\Gamma\rangle$, but not its distribution. This approximation
is certainly valid in the case of a single open channel in each lead,
the case for which we specialize our analytical and numerical
results. When the constrictions connecting leads to the cavity are
smaller than one electron wavelength but the cross section in the
leads is not, strong correlations among partial decay amplitudes at a
fixed resonance exist.\cite{Mucciolo95a,Alhassid95} For this general
situation it is still possible to incorporate different penetration
factors and the effect of correlations into our results, although we
do not believe that the final expressions could be cast in a simple
analytical form.  Nevertheless, one would need to obtain all
parameters either from the experiment, or from a satisfactory
dynamical model that comprises both cavity and leads.

Finite-temperature averaging caused by the rounding of the Fermi
distribution can be done straightforwardly. In the limit
$\Gamma_{c\nu}\ll kT<\Delta$ the conductance peak corresponding to an
{\it on-resonance} measurement is given by\cite{Meir91,Beenakker91}
\begin{equation}
\label{eq:peakcond}
    \widetilde G_\nu = \frac{2e^2}{h} \left( \frac{\pi}{2kT} \right)
	  g_\nu  \quad \mbox{with} \quad g_\nu =
	  \frac{\Gamma_{L\nu} \,\Gamma_{R\nu}} {\Gamma_{L\nu} +
	  \Gamma_{R\nu}} \; ,
\end{equation}
where $\Gamma_{L(R)\nu}$ is the decay width for the resonance $\nu$
into open channels in the $L(R)$ lead. In other words,
$\Gamma_{L(R)\nu} = \sum_{c\in L(R)} \Gamma_{c\nu}$. Equation
(\ref{eq:peakcond}) does not take into account thermal activation
effects. However, one should have in mind that in the tunneling regime
very small values of the partial decay widths may occur relatively
close in energy to large ones. In addition to that, the
single-particle spacing fluctuates and can reach values smaller than
$\Delta$. As a result, even at low temperatures (less than 100~mK) a
very short peak may be substantially influenced by neighboring
resonances. This effect causes the conductance peak to cross over to a
different regime, where the height becomes proportional to $T$ rather
than $1/T$ (even though the inequality $\Gamma_{R,L}\ll kT \ll\Delta$
is still approximately satisfied).\cite{Kleintip} We will avoid
further consideration of this or any other similar effect by
restricting our analysis to peaks whose heights have exclusively the
characteristic $1/T$ behavior.

In what follows we are going to evaluate the autocorrelator of
conductance peak heights in terms of an external parameter $X$ which
acts on the system Hamiltonian. (At this point we do not need to
specify $X$, which could be, for instance, a shape variable or a
magnetic flux.) In order to obtain the autocorrelator we use a
statistical approach that shall be described in detail in the next
subsection.  Unfortunately, a complete analytical solution for this
problem is not yet available and presently seems to be a formidable
task. Therefore our analytical treatment is based on a perturbative
expansion for small values of $X$. We resort to numerical
diagonalizations of large random matrices to conjecture the complete
form of the autocorrelator for all other values of $X$.

We also derive an exact analytical expression for the joint
probability distribution of conductance peak heights and their first
parametric derivative. This distribution, besides being of theoretical
interest by itself, can also be employed alternatively in the
calculation of the small-$X$ asymptotic limit of the conductance peak
correlator.

Before proceeding, we first present some basic quantities entering the
calculation. Since we are only considering two-point functions, the
knowledge of the partial decay widths at two parameter space points,
$X_1$ and $X_2$, is required. We will need to specify
$\Gamma_{c\nu}(X_2)$ only up to first-order terms in an expansion
around $\Gamma_{c\nu}(X_1)$, namely,
\begin{equation}
\Gamma_{c\nu}(X_2)=\Gamma_{c\nu}(X_1)+X\Lambda_{c\nu}(X_1)+O(X^2),
\label{eq:expgamma}
\end{equation}
where $X=X_2-X_1$. The coefficient $\Lambda$ is given by first-order
perturbation theory in terms of eigenfunctions, eigenvalues, and
matrix elements of the perturbation. Assuming that the Hamiltonian may
be written in the form $H(X)=H_0+XU$ and using Eq.~(\ref{eq:gamma}),
we find that
\begin{equation}
\Lambda_{c \nu} =
\sum_{\mu\ne\nu}\frac{\gamma_{c\nu}^\ast\gamma_{c\mu} U_{\mu\nu} +
\gamma_{c\nu}\gamma_{c\mu}^\ast U_{\nu\mu}}
{\varepsilon_\nu-\varepsilon_\mu},
\label{eq:Lambda}
\end{equation}
with $H_0\psi_\nu(\bbox{r})=\varepsilon_\nu\psi_\nu(\bbox{r})$ and all
partial decay amplitudes, eigenvalues, and matrix elements evaluated
at $X_1$. So far no statistical assumption has been made.

\subsection{Parametric conductance peaks autocorrelation function }
\label{sec:subcondpeakcor}

The starting point of our discussion of the conductance peak
autocorrelation function is Eq.~(\ref{eq:peakcond}). In this
subsection we shall mainly treat the situation of TR broken by an
external magnetic field. We identify the magnetic flux with the
previously introduced variable $X$. In the spirit of
RMT,\cite{Mehta91} the analysis that follows will lead to a universal
curve for the conductance peak autocorrelation, which is expressed in
terms of a scaled $X$ and the coupling of the dot to the external
leads, denoted by the parameters $a_{R,L}$. Both $a_{R,L}$ and the
typical scale of $X$ are system-specific quantities, depending on the
dot shape, the quality of the dot-lead couplings, and the Fermi
energy. Ideally, the scaling of $X$ should be extracted from the
experimental resonance (level) velocity correlator.\cite{Szafer93}
When this information is not available, it is not straightforward to
put $X$ in correspondence with experimental parameters. To overcome
this problem we invoke a dynamical system and shall discuss it at
length in Sec. III. Similarly, to obtain $a_{R,L}$ one needs to model
specific features of the tunneling barriers. We avoid this procedure
by expressing our results in terms of the mean decay width
$\langle\Gamma\rangle$, which then becomes a fitting parameter in
comparisons with experiments.

The autocorrelator of peak heights is defined as
\begin{equation}
C_g(X) \equiv \left\langle g_\nu\!\left( \bar{X}-\frac{X}{2} \right)
\, g_\nu\!\left( \bar{X}+\frac{X}{2} \right) \right\rangle -
\left\langle g_\nu\!\left( \bar{X}-\frac{X}{2} \right) \right\rangle
\left\langle g_\nu\!\left( \bar{X}+\frac{X}{2} \right ) \right\rangle
\;,
\label{eq:defCX}
\end{equation}
where $\langle \cdots \rangle$ denotes averages over the resonances
$\nu$ and over different values of $\bar{X}$, but will later be
interpreted as an average over the Gaussian unitary ensemble (GUE) of
random Hamiltonians.\cite{Mehta91} The difficulties involved in the
exact calculation of $C_g(X)$ become clear once we expand the
conductance peak height in powers of $X$ and assume translation
invariance (independence of $\bar{X}$),
\begin{equation}
C_g(X) = \sum_{n=0}^\infty \frac{(-1)^n X^{2n}}{(2n)!} \left\langle
\left[ \frac{d^n g_\nu(\bar{X})}{d\bar{X}^n} \right]^2 \right\rangle -
\langle g_\nu(\bar{X}) \rangle^2 .
\label{eq:expCX}
\end{equation}
(Notice that odd powers disappear since the autocorrelator is by
construction an even function of $X$.) As presented in
Eq.~(\ref{eq:expCX}), the task of finding the complete functional form
of $C_g(X)$ requires not just the knowledge of the second moment of
all derivatives of $g_\nu(\bar{X})$, but also the summation of the
Taylor series. Presently we do not know of any other alternative
method that could lead to the {\it complete} analytical form of
$C_g(X)$. An analogous situation is encountered in the calculation of
the more commonly studied level velocity autocorrelator\cite{Szafer93}
(see the detailed discussion of Ref.~\onlinecite{Simons93a}). The
small $X$ asymptotic limit can, however, be evaluated exactly through
standard random matrix methods. Moreover, an approximate curve for all
values of $X$ can be found through numerical simulations.

We start with the small-$X$ asymptotics. For this purpose we keep in
Eq.~(\ref{eq:expCX}) only the two lowest order terms,
\begin{equation}
C_g(X) = C_g(0) + C_g^{\prime\prime}(0) \frac{X^2}{2} + O(X^4) \;,
\label{eq:secndCX}
\end{equation}
with
\begin{equation}
C_g(0) = \left\langle \left( \frac{\Gamma_{R\nu}\,\Gamma_{L\nu}}
{\Gamma_{R\nu} + \Gamma_{L\nu}} \right)^2 \right\rangle - \left\langle
\frac{\Gamma_{R\nu}\,\Gamma_{L\nu}} {\Gamma_{R\nu} + \Gamma_{L\nu}}
\right\rangle^2
\label{eq:C0}
\end{equation}
and
\begin{eqnarray}
C_g^{\prime\prime}(0) & = & - \left\langle
\left(\frac{\Gamma_{R\nu}\Gamma_{L\nu}} {\Gamma_{R\nu}+\Gamma_{L\nu}}
\right)^2 \left( \frac{\Lambda_{R\nu}}{\Gamma_{R\nu}} +
\frac{\Lambda_{L\nu}}{\Gamma_{L\nu}} -
\frac{\Lambda_{R\nu}+\Lambda_{L\nu}}{\Gamma_{R\nu}+\Gamma_{L\nu}}
\right)^2 \right\rangle\;.
\label{eq:C0pp}
\end{eqnarray}
Now we introduce a statistical model: we assume that due to shape
irregularities and the presence of a magnetic field, the Hamiltonian
can be modeled as a member of the GUE.\cite{Mehta91} Under this
assumption, $C_g(0)$ has already been
obtained.\cite{Jalabert92,Bruus94,Mucciolo95a,Alhassid95} For example,
in the simplest case, where one has two equivalent uncorrelated open
channels (one in each lead), $C_g(0) =
4\langle\Gamma\rangle^2/45$. The second-order coefficient requires a
more complex treatment, since Eq.~(\ref{eq:C0pp}) involves not simply
eigenfunctions but eigenvalues as well. For Gaussian ensembles, a
great simplification is possible due to the statistical independence
of matrix elements of the Hamiltonian. This will allow us to divide
the ensemble average into averages over eigenfunctions and
eigenvalues.

For the sake of simplicity, we will take the matrix elements of $U$ as
Gaussian distributed,\cite{Uaverage} with
\begin{equation}
\langle U_{\nu\mu}\rangle = 0
\quad \mbox{and} \quad
\langle U_{\nu\mu}U_{\lambda\rho}\rangle = 
      \frac{\sigma^2}{N}\delta_{\nu\rho}\delta_{\mu\lambda} \;,
\label{eq:Uaverage}
\end{equation}
where $\sigma$ is a measure of the perturbation strength and $N$ is
the number of eigenstates (resonances) considered. It immediately
follows that
\begin{equation}
\langle\Lambda_{L(R)\nu}^2\rangle_{U} = \frac{2\sigma^2}{N}\sum_{c\in
   L(R)}\sum_{\mu\ne\nu} \frac{|\gamma_{c\mu}|^2|\gamma_{c\nu}|^2}
   {\varepsilon_{\nu\mu}^2}
\label{eq:averLambda1}
\end{equation}
and
\begin{equation}
\langle\Lambda_{R\nu}\Lambda_{L\nu}\rangle_{U} = \frac{\sigma^2}{N}
  \sum_{c\in R,c'\in L}\sum_{\mu\ne\nu} \left(
  \frac{\gamma_{c\nu}^\ast\gamma_{c'\nu}\gamma_{c\mu}\gamma_{c'\mu}^\ast}
  {\varepsilon_{\nu\mu}^2} + \mbox{c.c.} \right) \; ,
\label{eq:averLambda2}
\end{equation}
with $\varepsilon_{\nu\mu} = \varepsilon_\nu - \varepsilon_\mu$.  Up
to this point the formulation of the problem is very generic and no
restriction has been made either on the number of open channels or on
their transmission coefficients. A quick inspection of
Eqs.~(\ref{eq:averLambda1}) and (\ref{eq:averLambda2}) indicates that
the solution of the most generic case (multichannel leads) is
possible, but the algebra is very tedious and intricate. Recent
experimental realizations of quantum dots in the tunneling regime
\cite{Klein95,Folk95} indicate that one typically has {\it only one
open channel} in $L$ and $R$. Owing to that, we limit our calculations
to the one-channel case.

Carrying out the average over $U$ by using Eqs.~(\ref{eq:averLambda1})
and (\ref{eq:averLambda2}) to simplify Eq.~(\ref{eq:C0pp}), we get
\begin{equation}
C_g^{\prime\prime}(0) = - \left(\frac{2\sigma^2}{N}\right)
   \left\langle\left(
   \frac{\Gamma_{R\nu}\Gamma_{L\nu}}{\Gamma_{R\nu}+\Gamma_{L\nu}}
   \right)^4 \sum_{\mu\ne\nu}\frac{1}{\varepsilon_{\nu\mu}^2} \left[
   \frac{\Gamma_{R\mu}}{\Gamma_{R\nu}^3} +
   \frac{\Gamma_{L\mu}}{\Gamma_{L\nu}^3} + \left(
   \frac{\gamma_{R\nu}^\ast\gamma_{L\nu}\gamma_{R\mu}\gamma_{L\mu}^\ast}
   {\Gamma_{R\nu}^2\Gamma_{L\nu}^2} + c.c. \right) \right]
   \right\rangle_{\{\Gamma,\varepsilon_\mu\}}.
\label{eq:C0ppbis}
\end{equation}
As pointed out before, the independence of eigenvalues from
eigenfunctions allows the factorization of the right-hand side of
Eq.~(\ref{eq:C0ppbis}) into two decoupled averages, namely,
\begin{equation}
C_g^{\prime\prime}(0) = - \left( \frac{2\sigma^2}{N} \right)
A_{\Gamma} A_{\varepsilon} \;.
\label{eq:C0ppfac}
\end{equation}
The first factor is given by
\begin{eqnarray}
A_{\Gamma} & = & \left\langle\left(
\frac{\Gamma_{R\nu}\Gamma_{L\nu}}{\Gamma_{R\nu}+\Gamma_{L\nu}}
\right)^4 \left[ \frac{\Gamma_{R\mu}}{\Gamma_{R\nu}^3} +
\frac{\Gamma_{L\mu}}{\Gamma_{L\nu}^3} \right]
\right\rangle_{\{\Gamma_R,\Gamma_L\}},
\label{eq:Bamp}
\end{eqnarray}
where the average $\{\Gamma\}$ is taken over both leads $L$ and $R$
and resonances $\nu$ and $\mu$ with $\nu\ne\mu$ (terms with an odd
number of distinct amplitudes drop out). The second factor is
\begin{equation}
A_{\varepsilon} = \left\langle \sum_{\mu\ne\nu}\frac{1}
{\varepsilon_{\nu\mu}^2} \right\rangle_{\{\varepsilon_\mu\}}
\label{eq:Beig}
\end{equation}
and it involves only an average over the eigenvalues. Since the lead
contacts are typically farther than one wavelength $\lambda$ apart,
the decay widths at $R$ and $L$ are assumed to fluctuate
independently. Hence, the calculation of $A_{\Gamma}$ requires the
convolution of four $\chi^2$ distributions with two degrees of
freedom,\cite{Porter65} $P_2(\Gamma)=\exp(-\Gamma/\langle
\Gamma\rangle)$ (notice that we are considering a single open channel
per contact and identical leads), and we obtain $A_{\Gamma} = 2\langle
\Gamma\rangle^2/15$.

The eigenvalue average in Eq.~(\ref{eq:Beig}) is more complicated to
carry out and needs some additional assumption. Because we are going
to take the thermodynamic limit ($N\rightarrow\infty$) later on,
effects due to variations in the density of states are
negligible. Consequently, we may place the reference eigenstate
$\varepsilon_\nu$ at the center of the spectrum and write
\begin{equation}
A_{\varepsilon} \approx \frac{1}{\rho(0)}\sum_{\nu=1}^N
\left\langle\sum_{\mu\ne\nu}\frac{\delta(\varepsilon_\nu)}
{\varepsilon_{\nu\mu}^2}\right\rangle_{\{\varepsilon_\mu\}},
\label{eq:Beig1}
\end{equation}
where $\rho(E)=\langle\sum_{\nu=1}^N\delta(\varepsilon_\nu-E)\rangle$
is the average density of states. Before proceeding to evaluate this
quantity in the RMT framework, let us first discuss an illustrative
and extreme situation, the picket fence spectrum with spacing
$\Delta$. In this limit the averaged sum simplifies to
$(2/\Delta^2)\sum_{n=1}^\infty n^{-2}$, which readily yields
$A_{\varepsilon} = \pi^2/3\Delta^2$. Because the Gaussian ensembles
hypothesis implies fluctuations in the spectrum, the actual value of
$A_{\varepsilon}$ for the GUE should be larger but of the same order
of magnitude.

The average over eigenvalues in Eq.~(\ref{eq:Beig1}) can be restated
as an average over a spectral determinant,
\begin{eqnarray}
A_{\varepsilon} & = & \frac{1}{\rho(0)Z_N} \int \!
d\varepsilon_1\cdots d\varepsilon_N \exp \Biggl( -\frac{N}{2\lambda^2}
\sum_{\nu=1}^N \varepsilon_\nu^2 + 2\sum_{\nu\ne\mu} \ln \left|
\varepsilon_{\nu\mu} \right| \Biggr) \sum_{\nu=1}^N \sum_{\mu\ne\nu}
\frac{\delta(\varepsilon_\nu)} {\varepsilon_{\nu\mu}^2} \nonumber \\ &
= & \frac{N}{\rho(0)} \frac{Z_{N-1}}{Z_N} \left( \frac{N-1} {N}
\right)^{(N^2-3)/2} \left\langle \det(H^2) \mbox{tr}(H^{-2})
\right\rangle_{H} \;,
\label{eq:Beig2}
\end{eqnarray}
where $H$ is an $(N-1)\times(N-1)$ GUE matrix and the normalization
constant is given by \cite{Mehta91}
\begin{eqnarray}
Z_N & = & \int \!d\varepsilon_1\cdots d\varepsilon_N\exp\left(
      -\frac{N}{2\lambda^2}\sum_{\nu=1}^N\varepsilon_\nu^2 +
      2\sum_{\nu\ne\mu}\ln|\varepsilon_{\nu\mu}|\right) \nonumber \\ &
      = &
      (2\pi)^{N/2}\left(\frac{\lambda^2}{N}\right)^{N^2/2}\prod_{k=0}^N
      k! \; .
\label{eq:normaliz}
\end{eqnarray}
[Notice that $\Delta=1/\rho(0)=\lambda/\pi N$ is the mean level
spacing at the center of the spectrum.] In the limit of $N \gg 1$
Eq.~(\ref{eq:Beig2}) can be evaluated by the fermionic method
\cite{Itzykson89} (see Appendix \ref{sec:appendixA} for details) and
one finds that
\begin{equation}
A_{\varepsilon} = \frac{2\pi^2}{3\Delta^2}\;.
\label{eq:Beig3}
\end{equation}
As expected, the actual value of $A_{\varepsilon}$ is larger, but
still in fair agreement with the picket fence estimate.

The next step is to rescale the perturbation $X$ to a dimensionless
form, in such a way that all system-dependent parameters are
eliminated. One possibility is to use $\langle (dg/dX)^2 \rangle$ as
the rescaling parameter. As will become clear later on, we do not find
this procedure very interesting from the physics viewpoint because
this quantity cannot be easily calculated given the underlying
dynamical system. We rather follow an idea originally proposed by
Szafer, Simons, and Altshuler: Recall that the perturbation strength
$\sigma$ (the nonuniversal scale in the above calculations) also
appears in the level velocity correlator
$C_v(X)$,\cite{Szafer93,Simons93a}
\begin{equation}
\label{CvXdef}
C_v(X) = \frac{1}{\Delta^2}\left\langle
\frac{d\varepsilon_\nu}{d\bar{X}}\!\left(\bar{X}-\frac{X}{2}\right)\, 
\frac{d\varepsilon_\nu}{d\bar{X}}\!\left(\bar{X}+\frac{X}{2}\right) 
\right\rangle
\end{equation}
when evaluated at $X=0$, namely,
\begin{equation}
C_v(0)=\frac{\langle [U_{\nu\nu}]^2\rangle_U}{\Delta^2} =
\frac{\sigma^2}{N\Delta^2} \;.
\label{eq:vrms}
\end{equation}
The statement implicit in the original works of
Refs. \onlinecite{Szafer93} and \onlinecite{Simons93a} is that the
quantity $\sqrt{C_v(0)}$ sets the scale for {\it any} averaged
parametric function $\langle f(X)\rangle$, provided that the system
dynamics is chaotic in the classical limit. In
Sec.~\ref{sec:dynamical} we will show that $C_v(0)$ can be obtained by
semiclassical arguments, once details of the confining geometry of the
dot are known. Therefore, in analogy to their analysis of level
velocity correlator, we apply the rescalings $x=X\sqrt{C_v(0)}$ and
$c_g(x)=C_g(X)/C_g(0)$ to arrive at the following universal
(dimensionless) form:
\begin{equation}
c_g(x) = 1 -  \pi^2 x^2 + O(x^4) \;,
\label{eq:CXsmall}
\end{equation}
valid for $x\ll 1$.

To extract the nonperturbative part of $c_g(x)$, as well as its $x\gg
1$ asymptotic limit, we relied on a numerical simulation. We performed
a series of exact diagonalizations of random matrices of the form
$H(X)=H_1\cos(X)+H_2\sin(X)$, with $H_1$ and $H_2$ denoting two
$500\times 500$ matrices drawn from the GUE. This model for the
parametric dependence is rather convenient for the simulations and
later data analysis because it does not make the level density depend
on $X$, nor does it cause the eigenvalues of $H(X)$ to drift with
$X$. It is helpful to think of $[H(X)]_{kl}$ as the matrix element of
the Hamiltonian in a discrete space representation. As a result, for
the one-channel lead case we may simply equal decay widths to the
renormalized ($\langle\Gamma\rangle=1$) wave function intensities at a
given point,
\begin{equation}
\Gamma_{k\nu} = N|\psi_\nu(k)|^2\;,
\label{eq:widthnum}
\end{equation}
where $k$ is the site number and $\nu$ the eigenstate label. In this
way, we were able to generate more than $10^5$ different configurations 
of the two-lead geometry, out of which only $2\times 10^4$ were used 
(we stress that wave functions taken at different $\nu$ or $k$ are 
statistically independent when the size of the matrix is large enough).

\begin{figure}
\setlength{\unitlength}{1mm}
\begin{picture}(150,120)(0,0)
\put(10,10){\epsfxsize=13cm\epsfbox{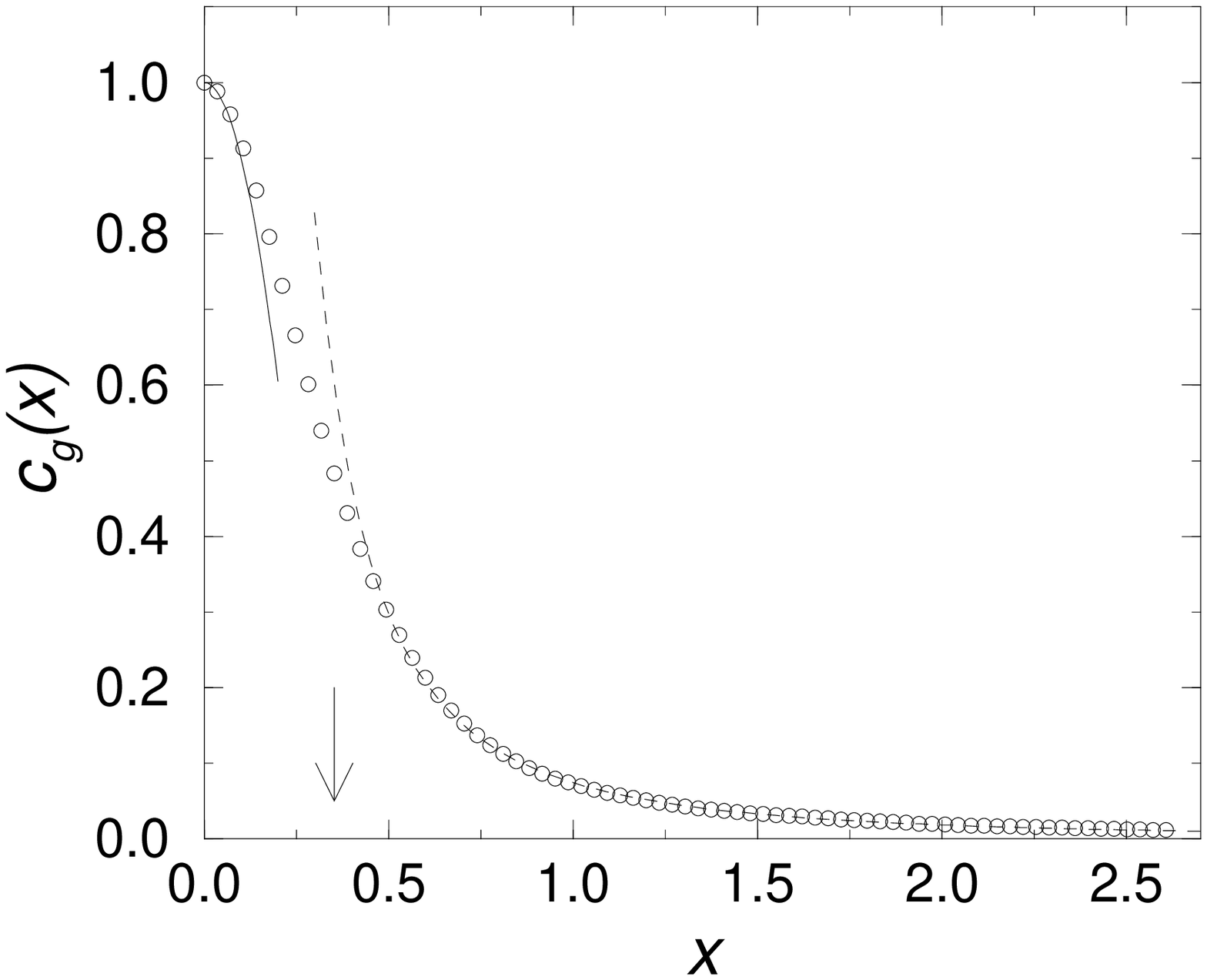}}
\put(48,40){\epsfxsize=8cm\epsfbox{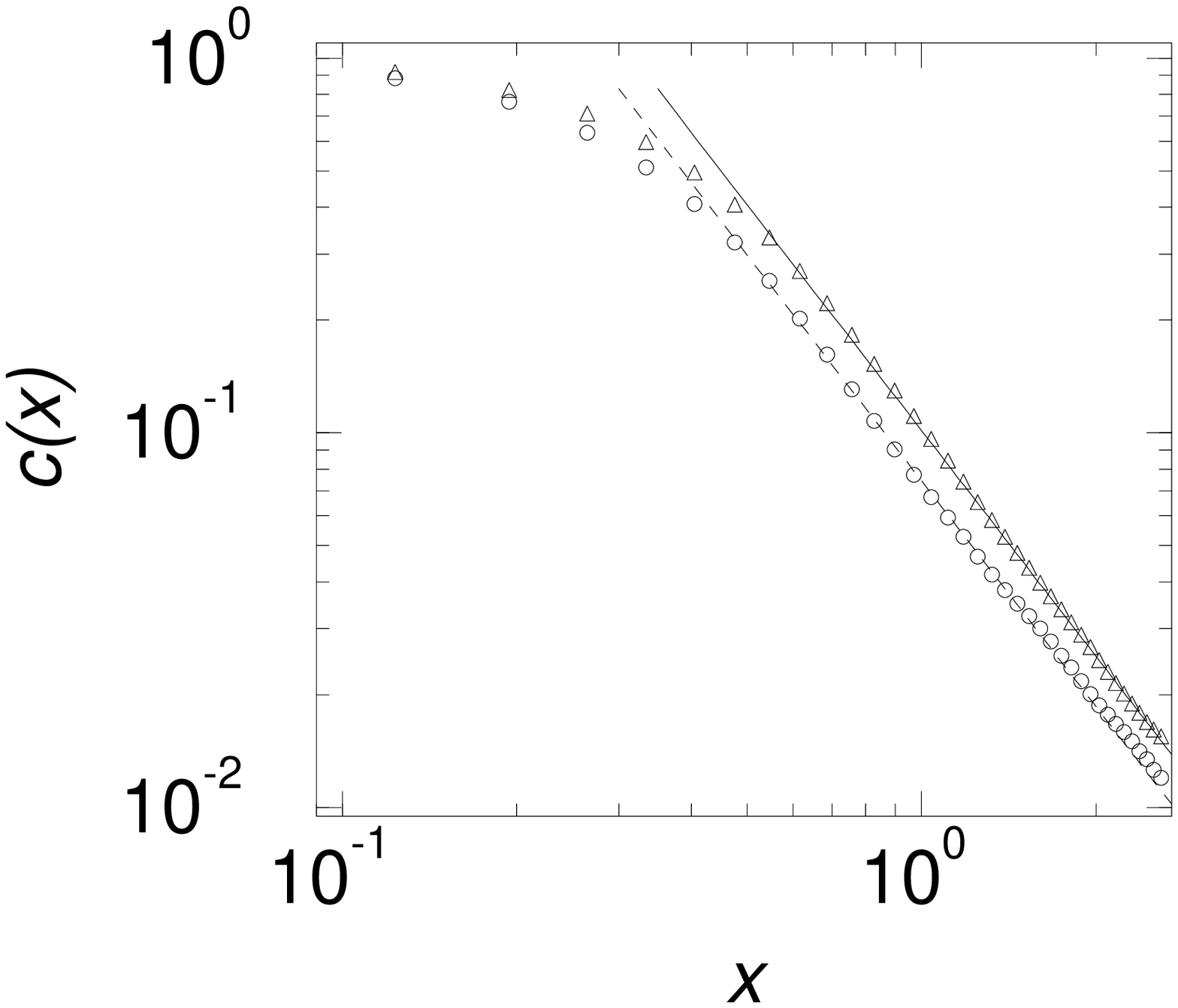}}
\end{picture}
\caption{The rescaled correlator of conductance peak heights
($\bigcirc$) obtained from the Hermitian random matrix simulations
(unitary ensemble). For comparison, the inset also shows the large $x$
asymptotics in the correlator of decay widths ($\bigtriangleup$). The
solid lines are the analytical predictions for the asymptotic
behaviors (when known) and the dashed line corresponds to the fitted
curve $c_g(x)=0.735(\pi x)^{-2}$. Statistical error bars are too small
to be seen. The arrow indicates the correlation width at half maximum
height.}
\label{fig:GUEsimul}
\end{figure}

For each realization of $H_1$ and $H_2$ we varied $X$ in the interval
$[0,\pi/8]$ and considered only the 100 central eigenstates in order
to avoid having to unfold the spectrum. In total we have run 50
realizations. The final result is presented in
Fig.~\ref{fig:GUEsimul}. For comparison, we have also shown in the
inset the result obtained for the decay width correlator,
\begin{equation}
C_\Gamma(X) =
\left\langle\Gamma_{k\nu}\!\left(\bar{X}-\frac{X}{2}\right)
\Gamma_{k\nu}\!\left(\bar{X}+\frac{{X}}{2}\right)\right\rangle -
\left\langle\Gamma_{k\nu}\!\left(\bar{X}-\frac{X}{2}\right)\right\rangle
\left\langle\Gamma_{k\nu}\!\left(\bar{X}+\frac{X}{2}\right)\right\rangle
\;.
\label{eq:corWW}
\end{equation}
We remark that this correlator is not directly accessible to
experiments in quantum dots (recall that conductance peak widths are
dominated by the thermal rounding of the Fermi
surface).\cite{Alhassid94} Here we have introduced it with the unique
purpose of checking the reliability of the numerical simulations in
the $x\gg 1$ range. Contrary to the situation for $c_g(x)$, both $x\ll
1$ {\it and} $x\gg 1$ asymptotic limits of the rescaled
$c_\Gamma(x)=C_\Gamma(X)/C_\Gamma(0)$ can be evaluated
analytically. One finds that (see Appendix \ref{sec:appendixB})
\begin{equation}
c_\Gamma(x) \longrightarrow \left\{\begin{array}{c} 1-2\pi^2 x^2/3
\quad \mbox{for} \quad x\ll 1 \\ 1/(\pi x)^2 \quad \mbox{for} \quad
x\gg 1
\end{array}\right. \;.
\label{eq:corWWasymp}
\end{equation}
The data obtained from the random matrix simulations indicate that the
asymptotic tail of $c_g(x)$ is well described by an $x^{-2}$ law. In
the light of Eq.~(\ref{eq:corWWasymp}), this is not very surprising:
If one lead were more strongly coupled to the cavity than the other,
say, $\Gamma_R\gg\Gamma_L$, we would have that
$c_g(x)=c_\Gamma(x)+O(\Gamma_L/\Gamma_R)$ and therefore $c_g(x\gg
1)\sim x^{-2}$ in leading order. We cannot rigorously prove, though,
that this asymptotic form is also exact when right and left leads are
identical.

\begin{figure}
\setlength{\unitlength}{1mm}
\begin{picture}(150,120)(0,0)
\put(10,10){\epsfxsize=13cm\epsfbox{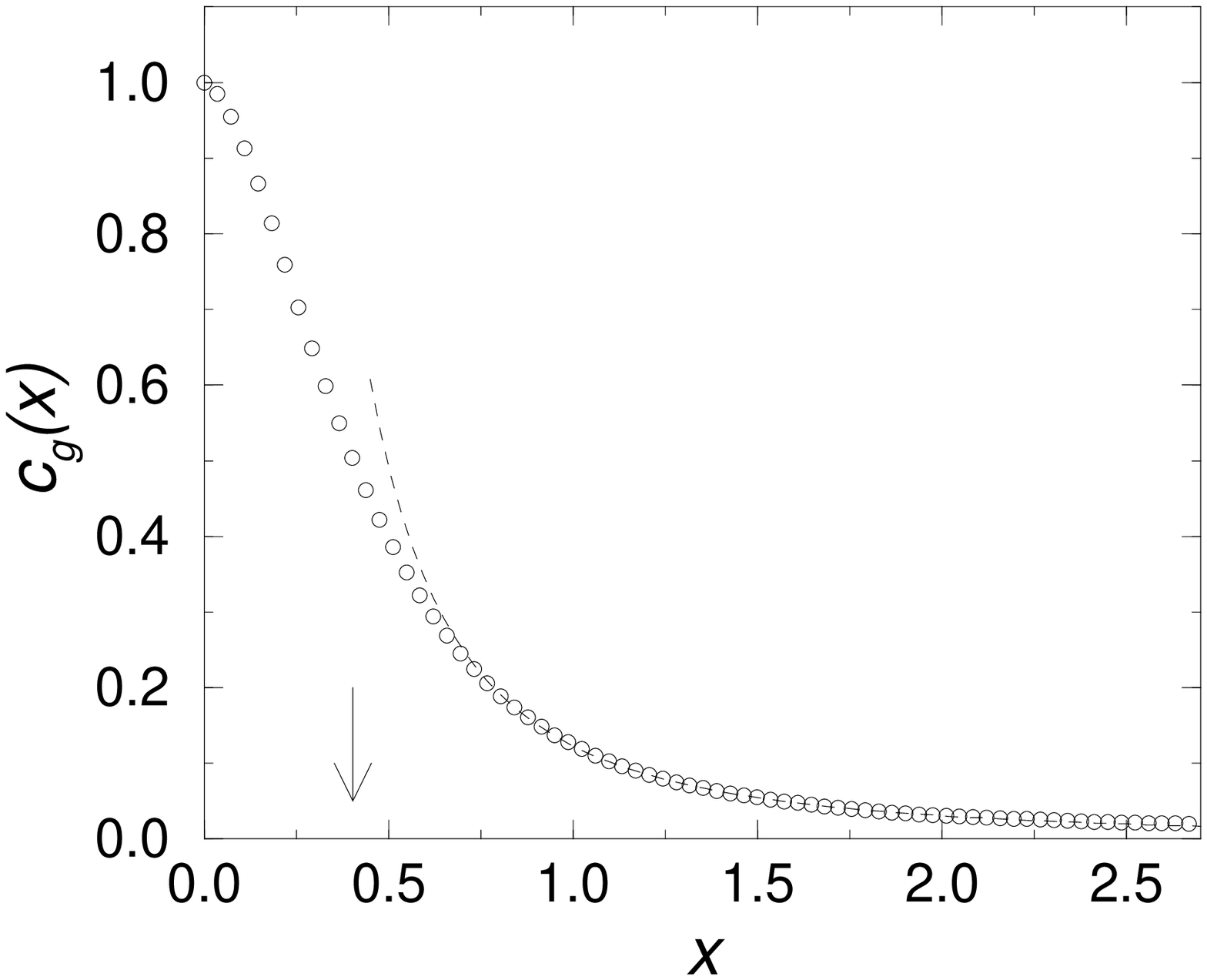}}
\put(48,40){\epsfxsize=8cm\epsfbox{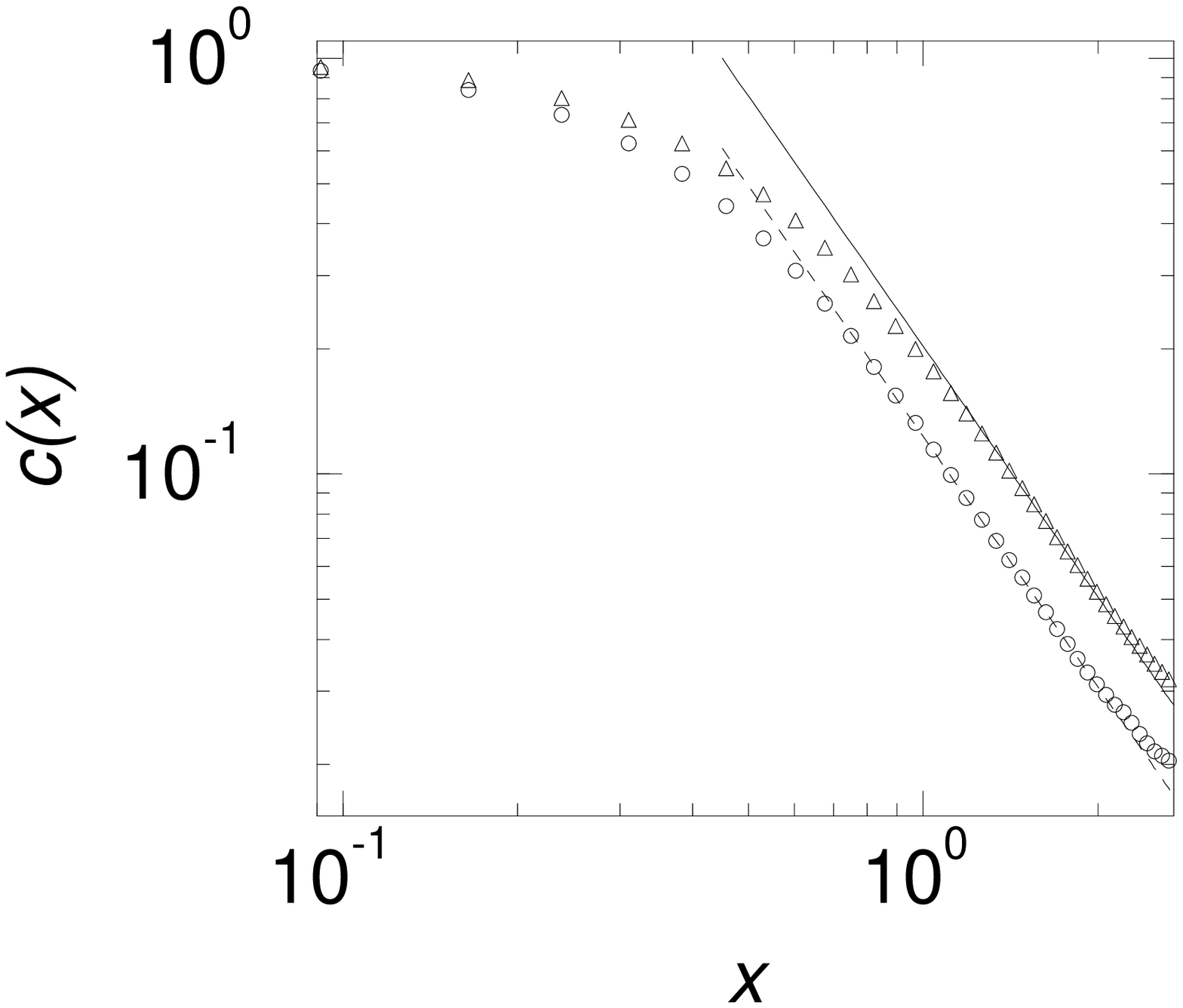}}
\end{picture}
\caption{The rescaled correlators of conductance peak heights and
decay widths from real random matrix simulations (orthogonal
ensemble), following the same conventions as
Fig.~\protect{\ref{fig:GUEsimul}}. The dashed line corresponds to the
fitting $c_g(x)=1.214(\pi x)^{-2}$ and the solid line is the
theoretical prediction for the $x\gg 1$ asymptotics of $c_\Gamma(x)$.}
\label{fig:GOEsimul}
\end{figure}

To conclude this subsection, we briefly discuss the universality class
of preserved TR without spin-orbit coupling (the case when spin-orbit
coupling is present, the symplectic ensemble, will not be discussed
since it is not relevant to semiconductor heterostructures). The
simplest experimental realization is a measurement of the evolution of
conductance peak heights as a function of shape deformation in the
absence of a magnetic field. The general approach is the same as above
and we assume that the system Hamiltonian can be modeled as a member
of the Gaussian orthogonal ensemble (GOE).\cite{Mehta91} However, the
$x\ll 1$ asymptotics of the conductance peak correlator is now more
difficult to calculate. This is not a daunting problem because, as
seen above, the numerical results reliably recover the correct
behavior of the correlator for small values of $x$. Consequently, for
the TR-preserved case we relied entirely on numerical simulations and
did not attempt any analytical calculation. We used the same
parametric dependence of $H(X)$, but this time drew $H_1$ and $H_2$
from the GOE. All other steps were identical to the GUE
simulation. The resulting correlation functions (after the proper
rescalings) are shown in Fig.~\ref{fig:GOEsimul}. Notice that the
large-$x$ asymptotics of both $c_g(x)$ and $c_\Gamma(x)$ are well
described by an $x^{-2}$ law, as for the GUE simulations. Here,
analogously to the GUE, we do not know how to prove analytically that
this decay is rigorously true for $c_g(x)$; on the other hand, we do
know that this power law decay is indeed exact for
$c_\Gamma(x)$.\cite{GOE} We emphasize that the most important
characteristic of the TR-preserved, universal $c_g(x)$ as compared to
the TR-broken one is the larger decay width. The widths at half
maximum height differ by approximately $20\%$.

\begin{figure}
\setlength{\unitlength}{1mm}
\begin{picture}(150,150)(0,0)
\put(10,10){\epsfxsize=13cm\epsfbox{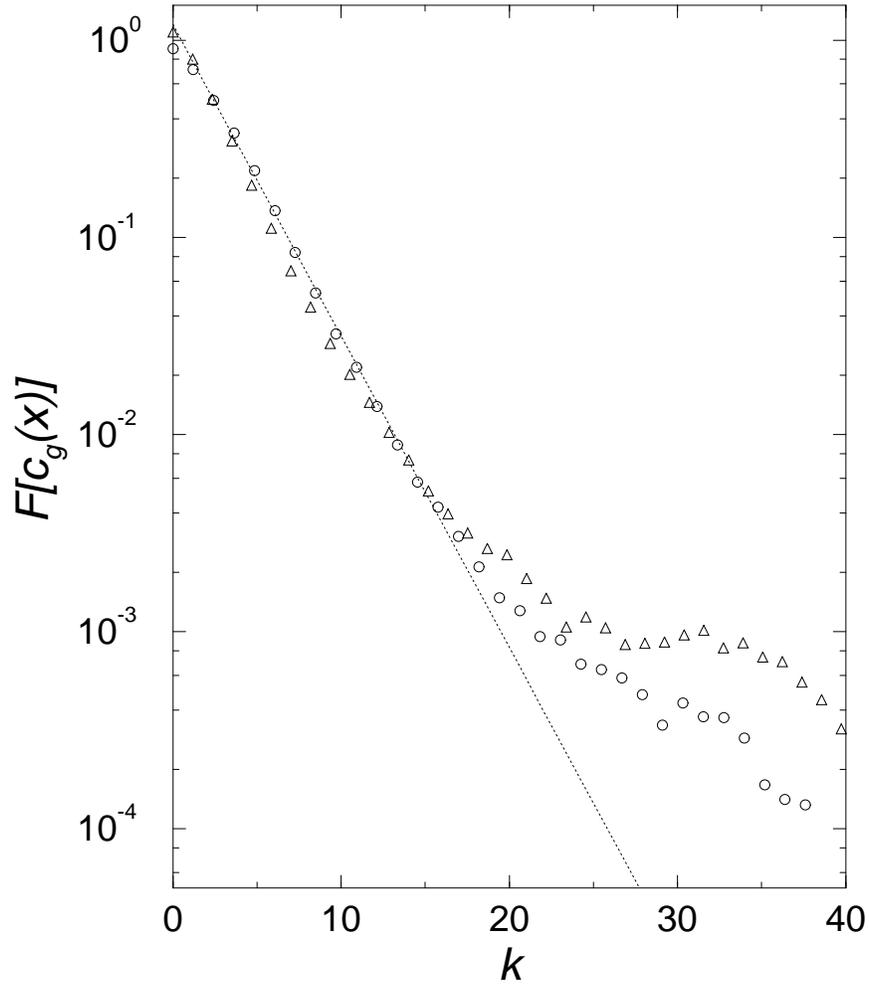}}
\end{picture}
\caption{The Fourier transform of the correlator of peak heights for
the GUE ($\bigcirc$) and GOE ($\bigtriangleup$) ensembles. The dashed
line is the curve $f(k)=0.125e^{-k/2.75}$ representing the Fourier
transform of a Lorentzian fitted to the GUE data.}
\label{fig:power}
\end{figure}

The power spectra of conductance peak height oscillations [the Fourier
transform of $c_g(x)$] for both GUE and GOE are shown in
Fig.~\ref{fig:power}. Notice that the behavior is exponential only
over a small range of $k$ values. One then concludes that a Lorentzian
can only be used as a {\it rough} approximation to the exact
curves. Besides, a Lorentzian cannot accommodate simultaneously the
small and large asymptotic limits of $c_g(x)$ presented above in
either TR-preserved or TR-broken cases.

\subsection{The joint distribution of conductance peak heights and 
their first parametric derivative (unitary ensemble)}
\label{sec:jointdistr} 

In this subsection we evaluate the joint probability distribution
\begin{equation}
Q(g,h) = \left\langle\delta\biglb( g - g_\nu(\bar{x}) \bigrb)
\delta\Biggl( h - \frac{dg_\nu(\bar{x})}{d\bar{x}}
\Biggr)\right\rangle \ ,
\label{eq:jointF1}
\end{equation}
where $\bar{x}=\sqrt{C_v(0)}\bar{X}$ and the Hamiltonian belongs to
the GUE. This distribution, although not suitable for direct
experimental investigations, allows one to easily evaluate higher
moments of the conductance peak height and its first parametric
derivative; in particular, one can use $Q(g,h)$ to obtain the two
first coefficients in the expansion of Eq.~(\ref{eq:secndCX}),
\begin{equation}
C_g(0) = \int_0^\infty \!dg \int_{-\infty}^{\infty} \!dh\ g^2 Q(g,h)
\label{eq:COident}
\end{equation}
and
\begin{equation}
C_g^{\prime\prime}(0) = - \frac{1}{C_v(0)} \int_0^\infty dg
\int_{-\infty}^{\infty} dh\ h^2 Q(g,h) \;.
\label{eq:C0ppident}
\end{equation}

We begin by recalling Eq.~(\ref{eq:Lambda}) and introducing the
coefficients $\Lambda_\nu$ into Eq.~(\ref{eq:jointF1}),
\begin{equation}
Q(g,h) = \left\langle\delta\Biggl(g -
\frac{\Gamma_{R\nu}\Gamma_{L\nu}}{\Gamma_{R\nu} +
\Gamma_{L\nu}}\Biggr) \delta(h - M_\nu )\right\rangle,
\label{eq:jointF2}
\end{equation}
where
\begin{equation}
M_\nu = \sqrt{C_v(0)}\left[ \frac{\Lambda_R}{\Gamma_{R\nu}} +
\frac{\Lambda_L}{\Gamma_{L\nu}} -
\frac{\Lambda_R+\Lambda_L}{\Gamma_{R\nu}+\Gamma_{L\nu}} \right]
\label{eq:Mexp}
\end{equation}
Taking the Fourier transform of Eq.~(\ref{eq:jointF2}) with respect to
$h$, we get
\begin{equation}
{\widetilde Q}(g,t) = \left\langle\delta\left(g -
\frac{\Gamma_{R\nu}\Gamma_{L\nu}}{\Gamma_{R\nu}+\Gamma_{L\nu}}\right)
\exp(itM_\nu)\right\rangle.
\label{eq:jointFfourier}
\end{equation}
 
Following the same assumptions of the previous subsection, we break up
the ensemble average into four partial averages. Recalling
Eqs.~(\ref{eq:averLambda1}) and (\ref{eq:averLambda2}), we first carry
out the average over the perturbation $U$:
\begin{equation}
\left\langle\exp(itM_\nu)\right\rangle_{U} =  \prod_{\mu\ne\nu}
\exp \Biggl( - \frac{g^4t^2\Delta^2}{\varepsilon_{\nu\mu}^2} \left|
\frac{\gamma_{R\nu}^\ast\gamma_{R\mu}}{\Gamma_{R\nu}^2} +
\frac{\gamma_{R\nu}^\ast\gamma_{L\mu}} {\Gamma_{L\nu}^2} \right|^2
\Biggr)\;.
\label{eq:apa17}
\end{equation}
Next, we evaluate the average over the decay widths
$\{\gamma_{R\mu},\gamma_{L\mu}\}$ for $\mu\neq\nu$ only,
\begin{equation}
\left\langle \exp\Biggl(-
\frac{g^4t^2\Delta^2}{\varepsilon_{\nu\mu}^2}
\left|\frac{\gamma_{R\nu}^\ast\gamma_{R\mu}}{\Gamma_{R\nu}^2}+
\frac{\gamma_{L\nu}^\ast\gamma_{L\mu}}{\Gamma_{L\nu}^2}\right|^2\Biggr)
\right\rangle_{\{\gamma_\mu\}} = \left[ 1 +
\frac{g^4t^2\Delta^2\langle\Gamma\rangle}{\varepsilon_{\nu\mu}^2}
\left(\frac{1}{\Gamma_{R\nu}^3} +
\frac{1}{\Gamma_{L\nu}^3}\right)\right]^{-1}\;.
\label{eq:jointF3}
\end{equation}
At this point, instead of also taking the average over the remaining
decay widths, we consider first the average over the eigenvalues
$\{\varepsilon_\nu\}$, $\nu=1,\cdots, N$. Here we use the following
relation, valid in the large-$N$ limit and for $\varepsilon_\nu$ at
the center of the spectrum:\cite{Mucciolo95b}
\begin{equation}
\left\langle\prod_{\mu\ne\nu}\left[1 +
\frac{(k\Delta/2\pi)^2}{\varepsilon_{\nu\mu}^2}
\right]^{-1}\right\rangle_{\{\varepsilon_\nu\}} = {\widetilde B}(k)\;,
\label{eq:eigaver}
\end{equation}
where ${\widetilde B}(k)$ is the Fourier transform of
\begin{equation}
B(s) = \frac{35+14s^2+3s^4}{12\pi(1+s^2)^4}\;.
\label{eq:B2}
\end{equation}
[The evaluation of the average in Eq.~(\ref{eq:eigaver}) requires a
generalization\cite{Andreev95} of the approach used to derive
Eq.~(\ref{eq:Beig3}). A brief description is given in Appendix
\ref{sec:appendixC}.] After inserting (\ref{eq:eigaver}) into
Eq.~(\ref{eq:jointF3}) and inverse Fourier transforming the result we
get
\begin{eqnarray}
Q(g,h) & = & \left\langle \delta\left(g -
\frac{\Gamma_{R\nu}\Gamma_{L\nu}}{\Gamma_{R\nu} +
\Gamma_{L\nu}}\right) \frac{1}{2\pi g^2\sqrt{\langle\Gamma\rangle}
\sqrt{\Gamma_{R\nu}^{-3} + \Gamma_{L\nu}^{-3}}}
\right. \nonumber \\ & & \times \left. B \Biggl(\frac{h}{2\pi
g^2 \sqrt{\langle\Gamma\rangle} \sqrt{\Gamma_{R\nu}^{-3} +
\Gamma_{L\nu}^{-3}}}
\Biggr)\right\rangle_{\{\Gamma_{R\nu},\Gamma_{L\nu}\}},
\label{eq:jointF4}
\end{eqnarray}
which is more conveniently expressed in the form
\begin{equation}
Q(g,h) = \frac{e^{-4g/\langle\Gamma\rangle}}{2\pi
\langle\Gamma\rangle^2 \sqrt{g/\langle\Gamma\rangle}}
H_{g/\langle\Gamma\rangle} \Biggl(\frac{h/\langle\Gamma\rangle}{2\pi
\sqrt{g/\langle\Gamma\rangle}}\Biggr)\;.
\label{eq:jointF5}
\end{equation}
The remaining average over the reduced widths
$\alpha_R=\Gamma_{R\nu}/\langle\Gamma\rangle$ and
$\alpha_L=\Gamma_{L\nu}/\langle\Gamma\rangle$ appears only in the
evaluation of the bell-shaped function
\begin{equation}
H_p(q) = \int_0^\infty d\alpha_R \int_0^\infty d\alpha_L\
{e}^{-(\alpha_R+\alpha_L-4p)} \ \frac{\delta\left(p -
\frac{\alpha_R\alpha_L}{\alpha_R+\alpha_L}
\right)}{\sqrt{p^3(\alpha_R^{-3} + \alpha_L^{-3})}}\ B
\Biggl(\frac{q}{\sqrt{p^3(\alpha_R^{-3} + \alpha_L^{-3})}}\Biggr)\;.
\label{eq:bellcurve1}
\end{equation}
In fact, one can easily carry out one of the above integrations and
arrive at
\begin{equation}
H_p(q) = 2p \int_0^\infty du\ e^{-pu} \frac{u+4}{\sqrt{u(u+1)}} B
\Biggl(q\sqrt{\frac{u+4}{u+1}}\Biggr)\;.
\label{eq:bellcurve2}
\end{equation}
It is possible to represent $H_p(q)$ in terms of special functions,
but we did not find it particularly clarifying and therefore do not do
it here. The asymptotic limits follow directly from the integral
representation shown above. For $q=0$, we obtain
\begin{equation}
H_p(0) \longrightarrow \left\{\begin{array}{c} (35/6\pi) \quad
\mbox{for} \quad p\rightarrow 0 \\ (70/3)\sqrt{p/\pi} \quad \mbox{for}
\quad p\rightarrow\infty\;, \end{array}\right. 
\label{eq:Hasympt1}
\end{equation}
whereas for a fixed $p\ne 0$, we have that
\begin{equation}
H_p(q) \sim O(q^{-4}) \quad \mbox{for} \quad q\gg 1 \;.
\label{eq:Hasympt2}
\end{equation}

Notice that Eqs.~(\ref{eq:jointF5}) and (\ref{eq:bellcurve2}) together
immediately lead to the same $x\ll 1$ asymptotics for the peak height
correlator shown in Eq.~(\ref{eq:CXsmall}). From
Eq.~(\ref{eq:Hasympt2}) it is obvious that $\langle h^n \rangle$
diverges for $n>2$. This can be ultimately related to the level
repulsion present in the spectrum: Strong anticrossings of levels can
cause anomalously large variations of the conductance peak height as a
function of the external parameter $X$. When this happens, one finds
that $h\sim 1/\omega$, where $\omega = |\varepsilon_{\nu+1} -
\varepsilon_\nu|/\Delta\ll 1$. Since the probability of this event
goes as $P(\omega)\sim\omega^2$ for the unitary
ensemble,\cite{Mehta91} one finds that $Q(g,h)\sim1/h^{4}$ for $h\gg
1$, in agreement with our exact calculation.

\section{Dynamical Model}
\label{sec:dynamical}

The aim of this section is to compare the results of the previous
section with exact numerical diagonalizations of a dynamical model.
The essential characteristic of a dynamical model for this type of
study is a fair resemblance to the actual experimental conditions,
combined with its adequacy to numerical computations. For this reason
we chose the (two-dimensional) conformal billiard penetrated by an
infinitely thin Aharonov-Bohm flux line carrying a flux of
$\phi$. This model was originally introduced in
Ref.~\onlinecite{Berry86} and later adopted in the study of
statistical features of conductance in quantum dots.\cite{Bruus94}
Using complex coordinates, the shape of the billiard in the $w$ plane
is given by $|z|=1$ in the following area preserving conformal
mapping:
\begin{equation} 
\label{w(z)}
w(z) = \frac{z + bz^2 + c e^{i \delta} z^3}{\sqrt{1+2b^2+3c^2}},
\end{equation}
where $b$, $c$, and $\delta$ are real parameters chosen in such a
manner that $|w'(z)| > 0$ for $|z| \leq 1$. The classical dynamics of
a particle bouncing inside this billiard is predominantly stochastic
and is unaffected by the presence of the flux line. To describe the
flux line we chose the following gauge for the vector potential
$\bf{A}$:
\begin{equation}
{\bf A}(w) = \frac{\phi}{2\pi} \left( \frac{\partial
f(w)}{\partial \mbox{Im}(w)}, - 
\frac{\partial f(w)}{\partial \mbox{Re}(w)}, 0 \right),
\end{equation}
where $ f(w) = \ln \left[ |z(w)| \right]$. This particular gauge,
combined with Neumann boundary conditions, permits a separation of the
Schr\"{o}dinger equation into polar coordinates $(r,\theta)$ of the
complex parameter $z=re^{i\theta}$. The eigenstates $\psi_{\nu}$ thus
obtained correspond to the resonant wave function appearing in
Eq.~(\ref{eq:gamma}). \cite{Bruus94} In our numerical treatment the
wave function $\chi_c$ in the lead is equal to a transverse sine
function multiplied by a longitudinal plane wave.
 
In practical calculations,\cite{Bruus94} one fixes the value of the
flux $\phi$ and uses as a truncated basis the lowest (in our case
1000) eigenstates of the circular billiard ($b=c=0$). These have the
form $J_{\nu}(\gamma_{n\nu}r) e^{il\theta}$, where $J_{\nu}$ is the
Bessel function of fractional order $\nu$ and $\gamma_{n\nu}$ is the
$n$th root of its derivative, $J'_{\nu}(\gamma_{n\nu}) = 0$.  The
dependence on the magnetic flux enters through $\nu =
|l-\phi/\phi_0|$, with $\phi_0$ as the flux quantum $h/e$. To solve
the Schr\"{o}dinger equation one has to calculate several thousand
matrix elements of the Jacobian ${\cal J} = |w'(z)|^2$ in this
$\phi$-dependent basis. This operation is very time consuming. Changes
in shape are not a major obstacle, since $b$, $c$, and $\delta$ act as
prefactors to the matrix elements and no further calculation is
necessary. (For instance, in Ref.~\onlinecite{Bruus94} only two
different values of $\phi$ needed to be used.) In the present work,
however, we wanted to change the flux to simulate the simplest
experimental setup and this required the use of 76 different values of
$\phi$.  The spectrum is seen to be not only $2\pi$ periodic in
$\phi$, but also symmetric around $\phi/\phi_0=1/2$. For this reason
and, furthermore, to avoid the special points 0 and 1/2 we let
$\phi/\phi_0$ vary in the interval $[0.1, 0.4]$. To circumvent the
overwhelming problem of too large amounts of computing time we
employed the following strategy. We calculated the Jacobian matrix
$\cal J$ for only seven different values of $\alpha= \phi/\phi_0$,
namely, from 0.10 to 0.40 in steps of 0.04. Then, for any other value
of $\alpha$, the matrix element ${\cal J}_{ij}(\alpha)$ was found by
polynomial interpolation. This was checked to give a relative error of
at most $10^{-7}$. To improve the statistics we also calculated the
eigenstates for five different shapes by keeping $b=c=0.2$ and letting
$\delta=k \pi/6$, with $k=1,2,3,4,5$. The spectra corresponding to
these shapes are statistically uncorrelated.\cite{Bruus94} Due to the
truncation of the basis only the lowest 300 of the calculated 1000
eigenstates were accurate enough to be used in the analysis. We also
discarded the lowest 100 eigenstates because of their markedly
nonuniversal behavior. We should emphasize that it is not a trivial
task to increase the number of usable states. For the asymmetric
conformal billiard no symmetry reduction of the resulting eigenvalue
problem is possible. The presence of the flux line constraints the
method of analysis to the diagonalization of large Hermitian matrices,
limiting the number of eigenstates that can be treated efficiently.

\subsection{Correlation functions for the billiard}
\label{sec:numerics}
 
In the following, we present the numerical results for the correlators
of level velocity, decay widths, and conductance peak heights. (Recall
that at present only the last can be directly measured in real
experiments.)
 
\begin{figure}
\setlength{\unitlength}{1mm}
\begin{picture}(150,150)(0,0)
\put(1,1){\epsfxsize=12cm\epsfbox{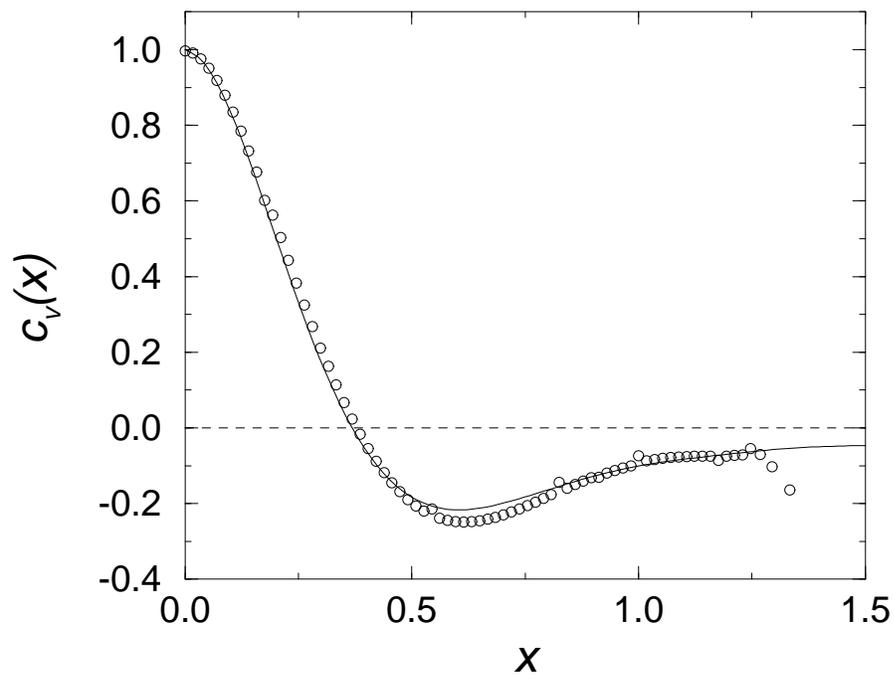}}
\end{picture}
\caption{The level velocity correlation function for the conformal
billiard ($\bigcirc$) averaged over five shapes, 76 values of the flux,
and 200 energy levels. The full line is the result of the GUE
simulation.}
\label{fig:cvBilliard}
\end{figure}

Figure~\ref{fig:cvBilliard} shows the level velocity correlator
defined by Eq.~(\ref{CvXdef}) rescaled according to $c_v(x) =
C_v(x)/C_v(0)$, with $x= \alpha/\sqrt{C_v(0)}$. The plotted data were
obtained by averaging over 76 equidistant values of $\alpha$ in the
interval $[0.1,0.4]$, over the eigenstates between 201 and 300, and
over the five different shapes mentioned above. We observe a good
agreement with the analytical results\cite{Simons93a} for small values
of $x$ and with random matrix simulations in general. A thorough
discussion of the scaling factor $\sqrt{C_v(0)}$ dependence on energy
and billiard shape is postponed to the next subsection.
 
\begin{figure}
\setlength{\unitlength}{1mm}
\begin{picture}(150,150)(0,0)
\put(1,1){\epsfxsize=12cm\epsfbox{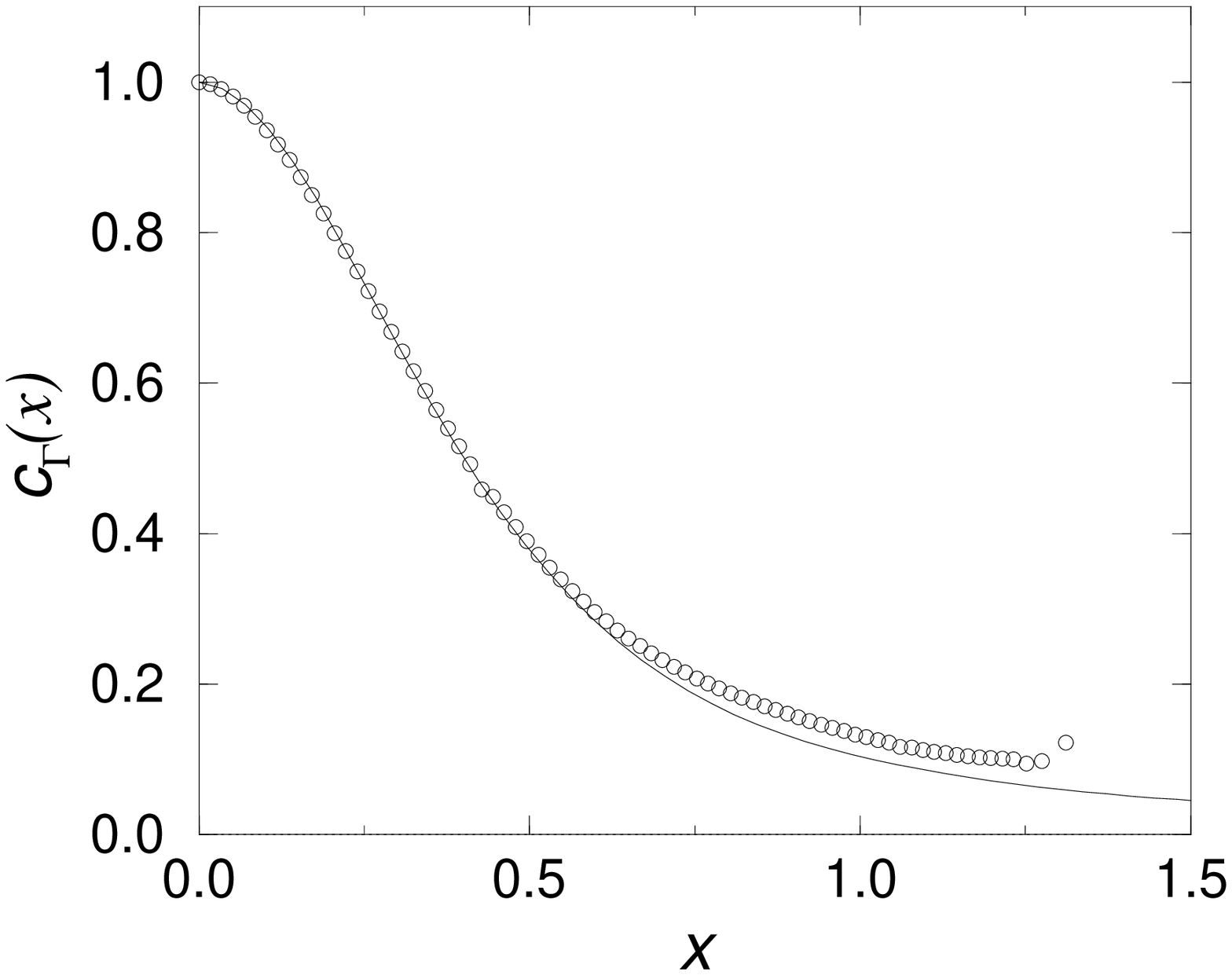}}
\end{picture}
\caption{The decay width correlation function for the conformal
billiard ($\bigcirc$) for a fixed shape ($b=c=0.2$ and $\delta=\pi/3$)
and averaged over 76 values of flux and 200 energy levels. The full
line is the result of the GUE simulation.}
\label{fig:cwBilliard}
\end{figure}

Next we employed the billiard model to obtain the decay width
autocorrelation function $c_{\Gamma}(x)$, Eq.~(\ref{eq:corWW}).  The
decay widths $\Gamma_{k\nu}$ for the billiard were calculated as
described in Sec.~\ref{sec:stochastic}. The position of the leads and
their widths were specified in the following way. Based on
autocorrelation calculations of the eigenfunction $\psi_{\nu}$ along
the perimeter of the billiard, we found that the spatial decorrelation
for the relevant levels (levels 201 to 300) takes place over a
distance around 1/60 of the perimeter. We therefore decided to take
the width of the leads to be 1/24 of the perimeter, i.e., 2.5 times
the decorrelation length, yielding 24 adjacent leads. To improve the
statistics we used all 24 lead position for each $\psi_{\nu}$. Due to
the relatively large width of the leads, adjacent leads are not
correlated, as was verified by obtaining the same result (with larger
fluctuations) using only every second or every third lead.  The result
of the calculation is shown in Fig.~\ref{fig:cwBilliard}. A fair
agreement with Eq.~(\ref{eq:corWWasymp}) is noted.
 
\begin{figure}
\setlength{\unitlength}{1mm}
\begin{picture}(150,150)(0,0)
\put(1,1){\epsfxsize=12cm\epsfbox{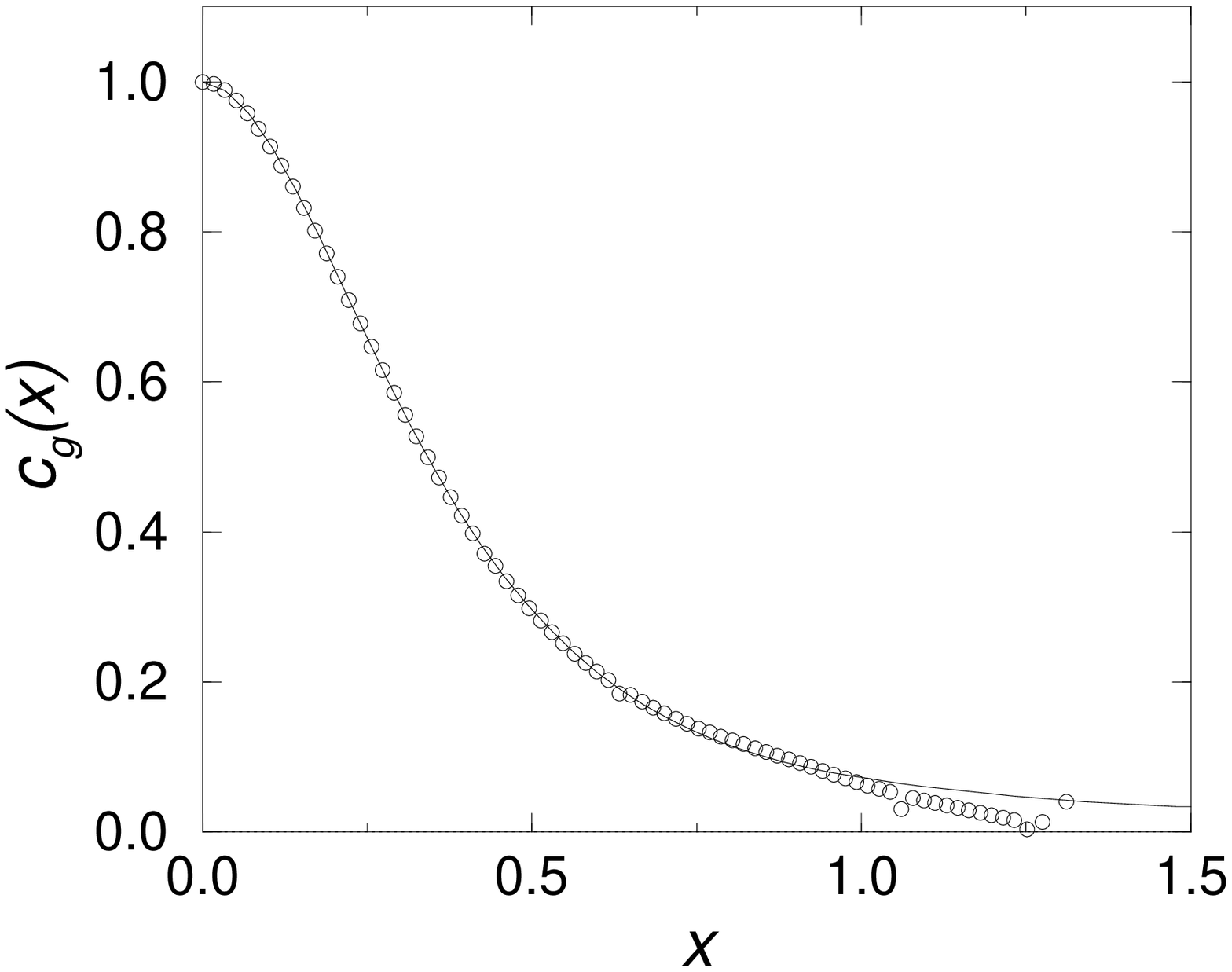}}
\end{picture}
\caption{The peak height correlation function for the conformal
billiard ($\bigcirc$) for a fixed shape ($b=c=0.2$ and $\delta=\pi/3$)
averaged over 76 values of flux and 200 energy levels. The full line
is the result of the GUE simulation.}
\label{fig:cgBilliard}
\end{figure}

Finally, we calculated the conductance peak correlation $c_g(x)$,
Eq.~(\ref{eq:defCX}), for the conformal billiard using the above
mentioned decay widths. We chose all possible pair configurations
among the 24 lead positions. The result is shown in
Fig.~\ref{fig:cgBilliard} and, again, a fair agreement with the
predictions of Sec.~\ref{sec:subcondpeakcor} is observed. Notice that
the data for the conformal billiard are not fully consistent with a
Lorentzian if we fix the $x\ll 1$ asymptotics of the curve to be
identical to Eq.~(\ref{eq:CXsmall}). A squared Lorentzian does not
seem to provide a better approximation either, although in a recent
experiment\cite{Folk95} $c_g(x)$ was measured and such a curve was
fitted to the data. It would be interesting to check how well our
result for $c_g(x)$ based on random matrix calculations matches the
available experimental data {\it without} any fitting parameter (in
Sec.~\ref{sec:energydep} we will present a way to predict the typical
field correlation scale).
 
For all three correlators we have noticed large statistical
fluctuations between data taken at different billiard shapes. We found
that most levels around the 300th one (the upper limit of reliability
in our calculation) still do not show more than one full oscillation
within the range of flux allowed by symmetry. The averaging over shape
deformation was thus crucial to get rid of the remaining nonuniversal
features. After averaging over the five values of $\delta$ mentioned
previously (see also the following subsection), we found that $C_v(0)$
near the $N$th level obeys the law $C_v(0) \approx 1.202 \sqrt{N}$.

We ascribe the small mismatch between theory and numerics,
particularly at $x>1$, to poor statistics. As mentioned before, the
only way to circumvent this problem is to compute higher
eigenstates. Fortunately, this difficulty does not appear in real
experiments where the magnetic flux is extended over the whole area of
the cavity and the dependence of $\phi$ is not periodic.

\subsection{Energy dependence of $C_v(0)$ for billiards}
\label{sec:energydep}

One of the important features of a dynamical model is that its quantum
fluctuations display a marked dependence on energy.  A {\it
quantitative} understanding of the field scale of the fluctuations and
its dependence on energy is important to put any random matrix result
in contact with measurements in quantum dots. For this purpose, the
semiclassical approach can be used in a relatively simple form.

The aim of this subsection is to discuss the energy dependence of the
level velocity correlator $C_v(0)$, which is a measure of the quantum
fluctuations, and show that one can successfully estimate this
quantity using exclusively classical quantities. It was already
properly noticed by Berry and Robnik \cite{Berry86} that the typical
flux necessary to induce a crossover from GOE- to GUE-like spectral
fluctuations in a chaotic billiard threaded by an Aharonov-Bohm flux
line scales with the energy as $E^{1/4}$. The origin of this
dependence has a simple semiclassical explanation which was nicely
worked out by Oz\'orio de Almeida and
coworkers.\cite{Almeida91,Bohigas95} The nonuniversal scaling $C_v(0)$
can be obtained in an analogous manner.

Our point of departure is a recent work by Berry and
Keating\cite{Berry94} based on the Gutzwiller trace formula. Most of
our semiclassical considerations follow their findings. However, our
interpretation and the method we use to quantify $C_v(0)$ are
different. To make the exposition self-contained, we shall briefly
describe points of their work which are relevant to our discussion and
comment when necessary.

The initial step is to approximate $C_v(\phi)$ by a two-point
correlator. (The nature of the approximation is evident, since to
track down the parametric evolution of a single eigenvalue for {\it
any } $\phi$ is a task that cannot be exactly achieved by considering
only Green's functions with a finite number of points.)  As in
Ref.~\onlinecite{Berry94}, we write
\begin{equation}
\label{ssacorr}
F_\eta(\phi, E) = \frac{1}{\delta\bar\phi} \int_{\delta\bar\phi}
   d\bar\phi \left\langle \frac{d}{d\bar\phi} N_\eta\!\left(\bar\phi
   -\frac{\phi}{2}\right) \frac{d}{d\bar\phi} N_\eta\!\left(\bar\phi
   +\frac{\phi}{2}\right) \right\rangle_{\delta E},
\end{equation}
where $N_\eta(E, \phi) = \sum_\nu \Theta_\eta \biglb( E -
\varepsilon_\nu(\phi)\bigrb)$ is a smoothed cumulative level density
which counts the number of single-particle states up to an energy $E$
at a magnetic flux $\phi$. For the smoothing it is convenient to adopt
the form $d\Theta_\eta(E)/dE \equiv {\rm Im}(E -i\eta)^{-1}/\pi$,
where the parameter $\eta$ is chosen to be much smaller than
$\Delta$. (We would like to point out a change in notation: Throughout
this subsection angular brackets $\langle \cdots \rangle$ will denote
energy averages in distinction to the previous ensemble averages.) The
energy average in Eq.~(\ref{ssacorr}) is taken over a range $\delta E$
around $E$.  Ordinarily, the average over the magnetic flux is taken
over a window $\delta\bar\phi$ in flux which corresponds to little
change in the classical dynamics but is semiclassically large, i.e.,
it corresponds to sizable fluctuations in the spectrum. For billiards
threaded by Aharonov-Bohm flux lines this is not an actual constraint,
since flux variations have no effect on classical
trajectories. Nonetheless, when considering the correlator of
Eq.~(\ref{ssacorr}) one has, in principle, to avoid $\bar\phi$
pertaining to the TR-breaking crossover regime.

When levels are much farther apart than $\eta$, it is straightforward
to show that $dN_\eta(\bar\phi)/d\bar\phi$ can be approximated by
$\Delta^{-1} d\varepsilon_\nu(\bar\phi)/d\bar\phi$. For $\phi$ larger
than a certain $\phi_c$ such that the correlation between level
velocities of different states $\mu \neq \nu$ is much weaker than for
$\mu =\nu$, $F_\eta(\phi, E)$ is equivalent to $C_v(\phi)$ and
independent of $\eta$, provided that the energy levels are taken to be
within $\delta E$. This equivalence also holds in the limit of
$\phi=0$ when the spectrum is nondegenerate. In summary,
$F_\eta(E,\phi)$ is a good approximation to the level velocity
correlator $C_v(\phi)$ only for $\phi=0$ and $\phi > \phi_c$. In fact,
$F_\eta(E,\phi)$ is also the quantity evaluated analytically by Szafer
and Altshuler \cite{Szafer93} in the context of disordered metallic
rings, using a diagrammatic perturbation theory based on impurity
averaging.

Following Ref.~\onlinecite{Berry94}, we now turn to the semiclassical
treatment. In particular, we specialize the results to billiards
threaded by a flux line. This simplifies the problem enormously since
if $S_n$ is the action of a periodic orbit $n$, upon applying a
magnetic flux $\phi$ we find that $S_n \rightarrow S_n+ 2\pi\hbar w_n
\phi/\phi_0$, where $w_n$ is the number of times the orbit $n$ winds
around the flux line. The cumulative level density is expressed
semiclassically using the Gutzwiller trace formula and one writes the
correlator as
\begin{equation}
\label{sccorr}
F_\eta(\phi, E) = \left( \frac{2\pi}{\phi_0} \right)^2
\left\langle \sum_{nm} |A_m A_n| \, w_n^2 \exp\left\{ i\left(
\frac{S_n-S_m}{\hbar} + 2\pi w_n \frac{\phi}{\phi_0} \right)
-\frac{\eta (T_n + T_m)}{\hbar} \right\} \delta_{w_n w_m}
\right\rangle_{\delta E},
\end{equation}
where the amplitudes $A_n(E)$ contain information on the stability of
the orbit $n$ as well as its Maslov index. The smoothing of the
staircase function gives rise to the exponential damping factor $\eta$
times $T_n(E)$, the period of the closed orbit $n$. In the
semiclassical limit, $N(E)\gg 1$, the flux range of the TR-breaking
crossover is much smaller than $\phi_0$. Therefore, without affecting
considerably the calculation for the pure TR-broken case, we take the
limit of $\delta\bar\phi \rightarrow \phi_0$.

Now one arrives at one of the delicate points of the semiclassical
approach. The correlator $F_\eta(\phi, E)$ is expressed as a sum of
diagonal and off-diagonal contributions. In contrast to the fact that
it is a settled matter how to compute the diagonal part, the
evaluation of the off-diagonal term is still an unsatisfactorily
solved problem. It seems that for $F_\eta$ it is
reasonable\cite{Berry94} to neglect the off-diagonal contribution and
we will do so hereafter.

The diagonal part of $F_\eta(\phi, E)$ in Eq.~(\ref{sccorr}) is still
difficult to evaluate since, in principle, it requires the knowledge
of the full set of periodic orbits up to the cutoff $\hbar/\eta$.
This can be simplified by the following considerations. For chaotic
systems, the number of periodic orbits grows exponentially as a
function of their length $L$ and ergodicity ensures that as the orbits
become longer they tend to explore the phase space more uniformly.
Thus, one can define a critical length $L_c$ corresponding to an
uniform coverage of the phase space by the periodic trajectories. For
a fixed energy $E$, $L_c$ determines $T_c$, the time when the Hannay
and Oz\'orio de Almeida sum rule \cite{Hannay84} is applicable,
allowing the sum over periodic orbits to be calculated as an integral
over orbital times. Applying this sum rule to the diagonal part of
Eq.~(\ref{sccorr}) one obtains
\begin{eqnarray}
\label{Fcomput}
F_\eta^{\rm{diag}}(\phi,E) & \approx & \frac{2}{\phi_0^2}
\int_{T_c}^\infty \frac{dT}{T}\; \overline{w^2(T) \cos\! \left( 2\pi
w(T) \frac{\phi}{\phi_0} \right)} \ \exp \left( -\frac{2\eta T}{\hbar}
\right)\;.
\end{eqnarray}
Here the overbar stands for an average over the phase space, or, in
practice, over an ensemble of trajectories. The replacement of the
summation over periodic orbits by an average over trajectories makes
the problem amenable for a computational evaluation of
$F_\eta^{\rm{diag}}(\phi,E)$. Such a procedure has already been
successfully used to estimate the GOE to GUE transition parameter in
the stadium billiard.\cite{Pluhar95}

\begin{figure}
\setlength{\unitlength}{1mm}
\begin{picture}(150,150)(0,0)
\put(1,1){\epsfxsize=12cm\epsfbox{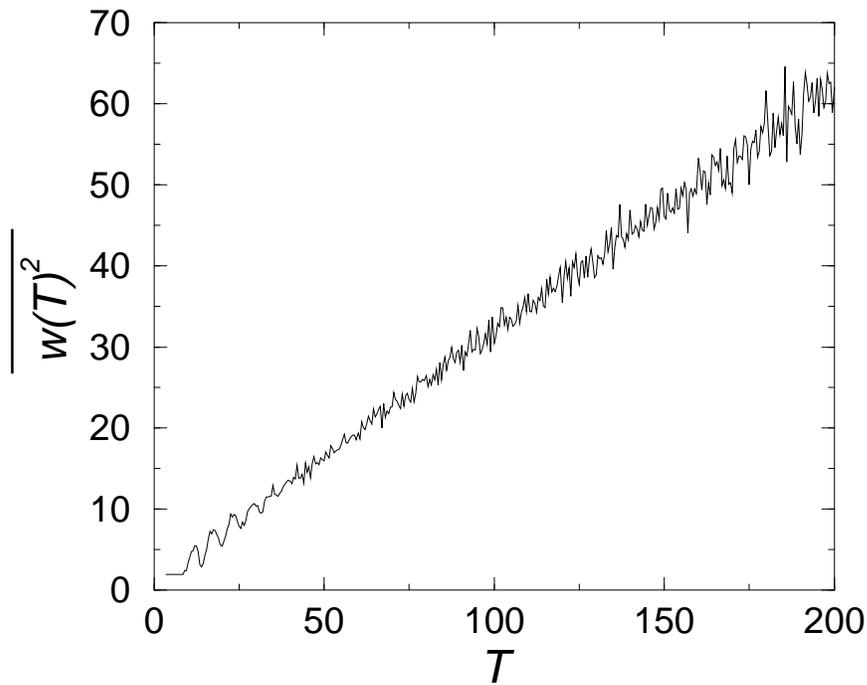}}
\end{picture}
\caption{Numerical estimate of the winding number variance
$\overline{w^2}$ as a function of orbit time $T$ in the classical
billiard with unit area, $v_F=1$, and $\delta=\pi/3$.}
\label{fig:wofT}
\end{figure}

\begin{figure}
\setlength{\unitlength}{1mm}
\begin{picture}(180,150)(0,0)
\put(1,1){\epsfxsize=12cm\epsfbox{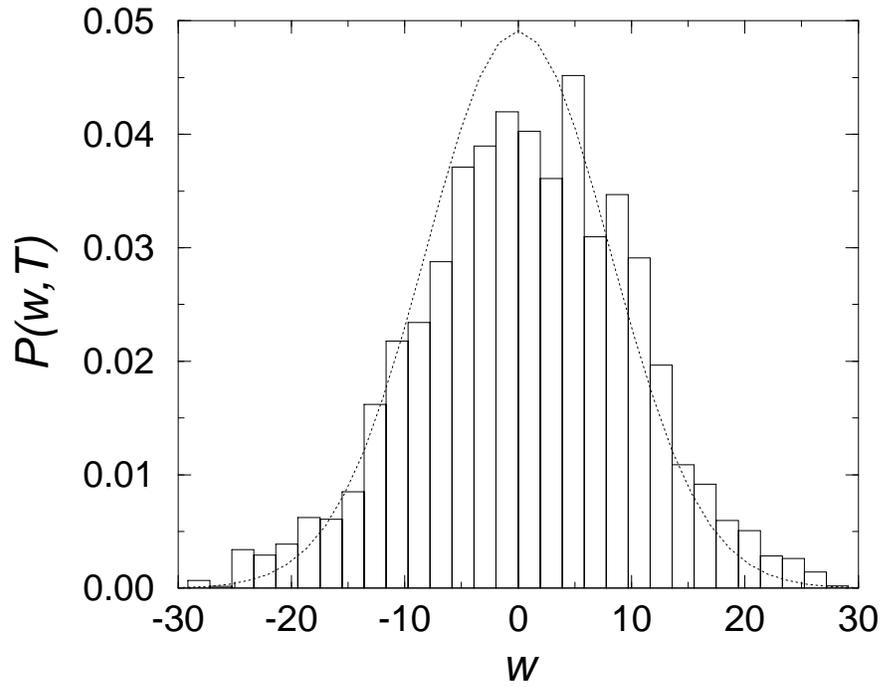}}
\end{picture}
\caption{The distribution of winding numbers (solid line) for $T=150$
in the classical billiard with unit area, $v_F=1$, and
$\delta=\pi/3$. The dotted line is a Gaussian curve with variance
given by Eq.~(\protect{\ref{eq:variance}}).}
\label{fig:Pofw}
\end{figure}

By writing the winding number as an integral over the angular
velocity, $w(T)=\int_0^T dt \,\dot\theta(t)/2\pi$, it is simple to see
that for an ergodic system $\overline{w(T)} = 0$. The winding number
variance is given by
\begin{eqnarray}
\label{wsimple}
\overline{w^2(T)} & = & \frac{1}{(2\pi)^2} \int_0^T dt \int_0^T
dt^\prime \, C(t^\prime - t) \nonumber \\ & \approx &
\frac{T}{(2\pi)^2} \int_0^\infty dt^\prime \, C(t^\prime) \;,
\end{eqnarray}
where $C(t)=\overline{\dot\theta(t^\prime)\dot\theta(t^\prime
+t)}$. For chaotic systems in general the correlator $C(t)$ decays
sufficiently fast in $T$ to assure the convergence of the integral in
Eq.~(\ref{wsimple}).\cite{Obs3} The knowledge of the winding number
distribution $P(w, T)$ allowed us to evaluate the phase space average
in Eq.~(\ref{Fcomput}). To determine $P(w, T)$ for the conformal
billiard we have randomly chosen $10^4$ initial conditions and
computed trajectories up to 250 bounces for the particular deformation
$b=c=0.2$ and different $\delta$'s. We have generated a histogram
recording $w$ and $T$ every time a trajectory winds around the flux
line located at the origin.\cite{Obs4} The results displayed in
Fig.~\ref{fig:wofT} confirm Eq.~(\ref{wsimple}). The variance of
$w(T)$ is better written as\cite{Pluhar95}
\begin{equation}
\overline{w^2(T)} = \kappa \left( \frac{2E}{m \cal{A}} \right)^{1/2} T
\;,
\label{eq:variance}
\end{equation}
where $\cal{A}$ is the billiard area and $\kappa$ is a {\it
system-dependent} quantity computed for the scaled billiard with unit
area and trajectories with unit velocity. Moreover, our numerical
results give us confidence that for any given time $P(w, T)$ is a
Gaussian distribution (see Fig.~\ref{fig:Pofw}). This is in agreement
with the conjecture of Ref.~\onlinecite{Berry86} for periodic orbits.

Substituting this result into (\ref{Fcomput}) and baring in mind that
in the semiclassical limit $\eta T_c/\hbar \ll 1$, for $\phi=0$ we
obtain
\begin{equation}
\label{Ffinal}
F_\eta^{\rm{diag}}(0,E) \approx \frac{1}{\phi_c^2} \;.
\end{equation}
Using the leading term in the Weyl formula, $N(E)\approx {\cal
A}mE/2\pi\hbar^2$, we can write
\begin{equation}
\label{eqscale}
\phi_c = \frac{\phi_0} {\left[ 4\pi\kappa^2 N(E) \right]^{1/4}} \;.
\end{equation}
Notice that $\phi_c\ll\phi_0$.

\begin{figure}
\setlength{\unitlength}{1mm}
\begin{picture}(180,150)(0,0)
\put(1,1){\epsfxsize=12cm\epsfbox{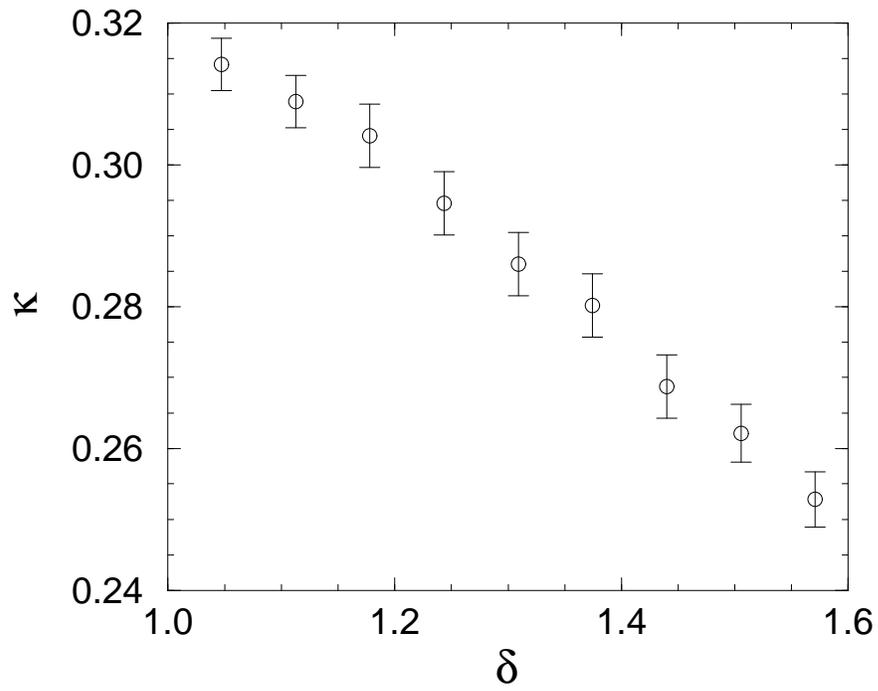}}
\end{picture}
\caption{The numerical coefficient $\kappa$ as a function of geometry
for the classical conformal billiard.}
\label{fig:fig9}
\end{figure}

\begin{figure}
\setlength{\unitlength}{1mm}
\begin{picture}(180,150)(0,0)
\put(1,1){\epsfxsize=12cm\epsfbox{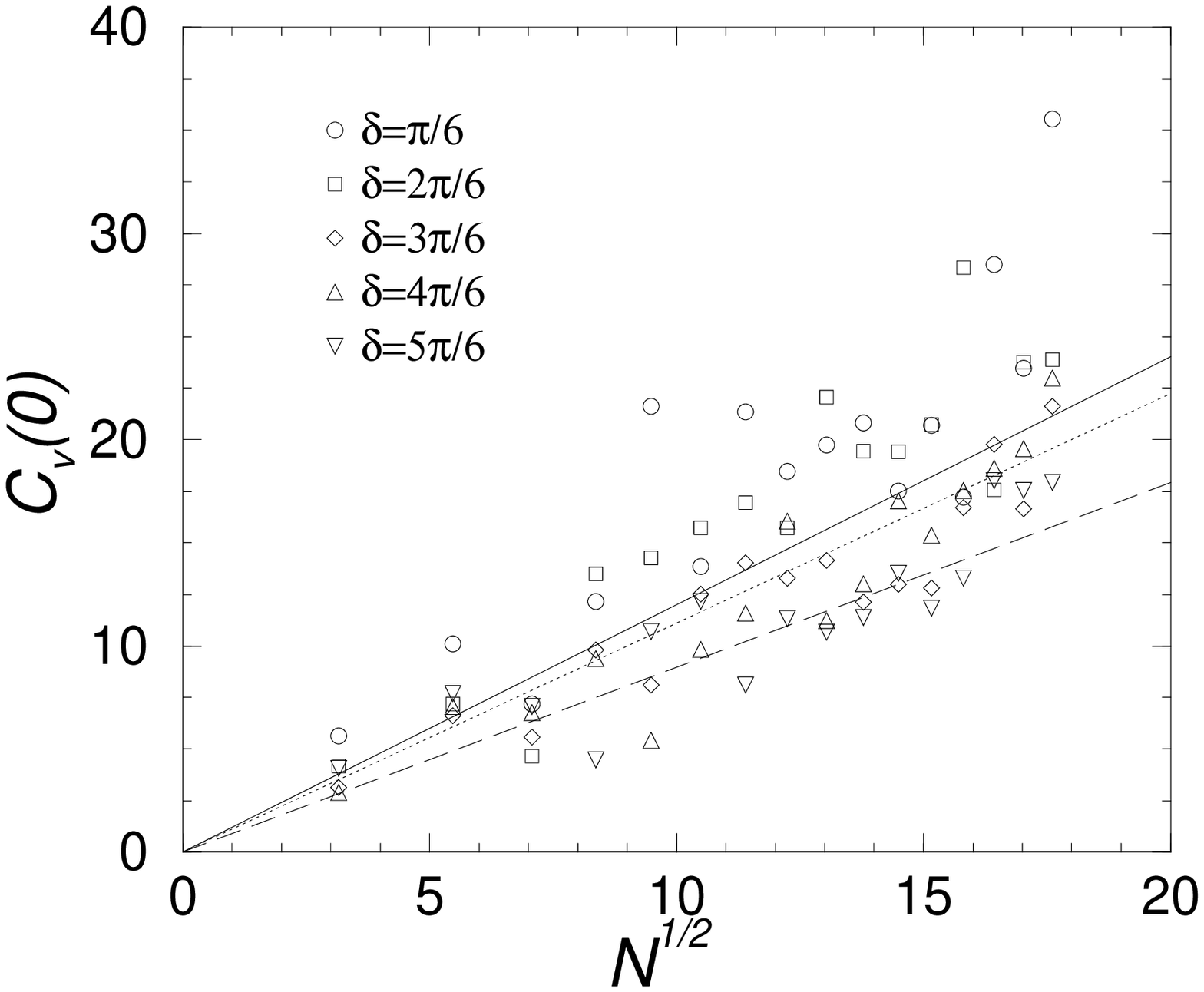}}
\end{picture}
\caption{$C_v(0)$ as a function of $\protect{\sqrt{N}}$ for the
conformal billiard, with $N$ as the eigenstate number. The symbols
indicate the data obtained from the exact numerical diagonalization
(fully quantum) for different geometries. The solid line is the total
average over all data: $C_v(0)=1.202\protect{\sqrt{N}}$. The dashed
and dotted lines are the semiclassical estimates
[Eq.~(\protect\ref{Ffinal})] for $\delta=\pi/2$ ($\kappa=0.253$) and
$\delta=\pi/3$ ($\kappa=0.314$), respectively.}
\label{fig:scC0}
\end{figure}

Although $F(\phi,E)$ fails as an accurate approximation for
$C_v(\phi)$ over the entire range of $\phi$, we expect it to work for
$\phi=0$ and $\phi\gg\phi_c$. Therefore, the semiclassical quantity
$\phi_c$ given above should yield a good approximation to the exact
inverse field scale $\sqrt{C_v(0)}$. We used the same procedure
described to obtain $P(w,T)$ to compute $\kappa$ as a function of
$\delta$. The results are show in
Fig.~\ref{fig:fig9}. Figure~\ref{fig:scC0} shows $C_v(0)$ as a
function of $N$ from quantum mechanical (Sec.~\ref{sec:numerics}) and
semiclassical ($\phi_c^{-2}$) calculations. From the numerical
diagonalizations we found that the proportionality factor between
$C_v(0)$ and $N^{1/2}$ varies between 0.94 ($\delta=5\pi/6$) and 1.54
($\delta=\pi/6$), while the semiclassical estimate gives 0.90 for
$\delta=\pi/2$, for instance. Clearly, both quantum and semiclassical
calculations indicate that $C_v(0)$ depends on the billiard
shape. When comparing the results of these two calculations, one
should also note the size of the large fluctuations in the data
presented in Fig.~\ref{fig:scC0}. As already remarked in the previous
subsection, these large statistical fluctuations are due to the
limited data set used in the simulations.

To put the semiclassical result in direct contact with experiments,
one needs some system-specific information to compute $\kappa$. In the
above discussion, $\kappa$ gives a measure of how fast the winding
number variance increases with time. For the more physical situation
of extended $B$ fields, $\kappa$ measures the rate of increase in the
variance of accumulated areas as a function of time.\cite{Almeida91}
It is interesting to notice that the semiclassical interpretation of
the scales of the conductance autocorrelation function for quantum
dots and open cavities is very similar. Since quantum dots have
$\Gamma\ll\Delta$, the escape time $\hbar/\Gamma$ is always
semiclassically large. In this regime, the physics is dominated by the
classical decorrelation time $\tau$ implied in
Eq.~(\ref{wsimple}). The situation is very different for open
cavities, where the escape time plays an important role, since it is
comparable with $\tau$. For the quarter stadium and extended $B$
fields, Ref.~\onlinecite{Pluhar95} gives $\kappa\approx 0.3$. In this
study we observed that $\kappa$ is a relatively robust number, since
very different shapes of the conformal billiard give values of
$\kappa$ that differ at most by $60\%$.  Therefore we believe that
with help of Eq.~(\ref{eqscale}) and taking $\kappa$ to be of order
unity, one can estimate the magnitude of $\phi_c$ for other chaotic
billiards.

\section{Conclusions and discussion}
\label{sec:conclusions}

In this paper, we have proposed that the universal form of the
parametric correlator of conductance peak heights indicates the
chaotic nature of the electron dynamics in quantum dots in the Coulomb
blockade regime. In experiments, the simplest parameter to vary is an
external magnetic field. Whereas random matrix theory provides the
universal form of the correlation function, the nonuniversal field
scale can be understood in simple semiclassical terms: it is related
to the average winding number per unit of time of periodic orbits
bouncing between the confining walls of the quantum dot. This field
scale is rather sensitive to the geometry of the dot and the Fermi
energy. To compare our analytical and numerical predictions against
the experimental result, the magnetic flux through the dot has to be
larger than one quantum unit of flux $h/e$, but such that the
cyclotron radius is much larger than the dot diameter. The former
condition assures that time-reversal symmetry is broken; the latter
implies that the bending of classical trajectories is mainly due to
scattering by the boundaries. We point out that electron-electron
correlations can be taken into account by assuming that the
single-particle spectrum results from a self-consistent (Hartree-Fock)
treatment.\cite{Obs5}

Our theoretical prediction for the correlator of peak heights is based
on the hypothesis that the statistical properties of the system
Hamiltonian can be described by random matrix theory. Although we
could only derive analytically expressions for the limit of small
field variations, the complete form of the correlation function was
obtained by numerical simulations of large Gaussian matrices. We have
compared the random matrix results with the exact correlator obtained
from the conformal billiard after averaging over energy and shape
deformation. The agreement found was good, given the limitations
imposed by the size of the data set. In addition to that, we found
that the result of the classical calculation for the magnitude of the
field scale and its dependence on energy matches moderately well the
quantum result.

For experimental tests of our theory, it is important to note that
dephasing in the small quantum dot has to be kept low enough, namely,
the dephasing length has to be larger than the system size. Also, the
parametric results presented for the GUE case were illustrated in a
dynamical model in which only the magnetic field was varied. This was
very convenient, since it allowed us to the relate the field scaling
parameter to the underlying classical dynamics using a well-developed
formalism. Alternatively, the external parameter could be taken as the
shape variation. For $B=0$, the conductance autocorrelation function
should then follow our numerical results for the GOE case. Or, by
varying the shape with $B>B_{\rm crossover}$, the conductance peak
correlator should be given by our GUE results. (While finishing this
work we learned that such an experiment has already been
performed.\cite{Folk95}) Unfortunately, the semiclassical analysis for
these situations is more difficult than the one presented here and
still remains an open problem.

Some billiard geometries have well-pronounced short periodic orbits
which, for insufficient averaging over energy and magnetic field, can
lead to strong nonuniversal features to the curves presented in this
work. We believe that overcoming this problem will be one of the most
serious challenges for the experiments. In particular, to verify
experimentally the asymptotic behavior of $c_g(x)$ should be very
difficult. Since $c_g(x)$ involves the subtraction of two numbers and
for $x\gg 1$ these numbers become very close, statistical
(nonuniversal) fluctuations can easily drive the experimental $c_g(x)$
below zero. Another cause of deviations from the predicted universal
behavior, presumably weaker, is the existence of wave function
correlations which extend over the dot. In other words, our findings
assume that the channels at different leads are decorrelated, which
may not be completely true if the dot size is not much larger than the
electron wavelength. The agreement of our random matrix results with
the numerical calculations using the conformal billiard supports the
assumption of independent channels for $N>200$. Smaller systems should
be more influenced by short or direct orbit effects. We should also
mention that if there were strong correlations between heights of
neighboring peaks of a given sequence, they would influence mostly the
$x\gg 1$ region of the correlator. This is because averaging over a
finite range of magnetic field usually yields less statistics for
large field differences and, consequently, more pronounced data
correlation effects.

Lastly, we point out the fact that, independently of previous
considerations, another very interesting experiment is a direct
measurement of $c_v(x)$. Despite the very large Coulomb energy, which
makes the conductance peak spacing very regular at first sight very, a
careful experiment should be able to observe the small fluctuations of
the peak position as a function of an applied magnetic field. This
will give direct information about the single-particle level
dispersion or, equivalently, $c_v(x)$. It will also provide a direct
test of our estimate of the flux correlation scale $\phi_c$.

{\it Note added.} After the submission of this manuscript we learned
of similar work by Alhassid {\it et al.}\cite{Alhassid96}
 
\acknowledgements

H.B.\ was supported by the European Commission under Grant No.\
ERBCHBGCT 930511. C.H.L.\ acknowledges the financial support of the
National Science Foundation. Both H.B.\ and C.H.L.\ thank NORDITA for
its hospitality. We are grateful to B. Altshuler, O. Klein, C. Marcus,
B. Simons, N. Taniguchi, and S. Tomsovic for helpful discussions and
to R. Jalabert for suggestions concerning the manuscript. We also
thank H. Baranger and C. Marcus for providing us with
Refs. \onlinecite{Chang95} and \onlinecite{Folk95}, respectively,
prior to publication.

\appendix
\section{Evaluation of $\left\langle\det(H^2)\,
    \lowercase{{\rm tr}}(H^{-2})\right\rangle_H$ }
\label{sec:appendixA}

This Appendix is devoted to a rather detailed evaluation of the
average over the determinant shown in Eq.~(\ref{eq:Beig2}). Let us
call
\begin{equation}
c_N = \langle\det(H^2)\, \mbox{tr}(H^{-2}) \rangle_{H}\; ,
\label{eq:apc1}
\end{equation}
where $H$ is now an $N\times N$ GUE matrix. First, we notice that
there is a more convenient way to express this average, namely,
\begin{equation}
c_N = \left. \frac{d\, f(a)}{da} \right|_{a=0},
\label{eq:apc2}
\end{equation}
where the generating function $f(a) = \langle \det (H^2+a\openone_N)
\rangle_H$ can be evaluated by the fermionic method:\cite{Itzykson89}
\begin{equation}
f(a) =
\langle\det(H+\alpha\openone_N)\det(H-\alpha\openone_N)\rangle_H =
\left\langle \int \! d[\chi] \exp[-\chi^\dagger(H\otimes\openone_2 +
\alpha \openone_N\otimes L)\chi] \right\rangle_H,
\label{eq:apc3}
\end{equation}
where $a = -\alpha^2$, $L=\mbox{diag}(1,-1)$, and $\chi^T =
(\chi_1^T,\chi_2^T)$, with $\chi_1$ and $\chi_2$ representing
$N$-component fermionic vectors. Averaging over the GUE matrix $H$ we
find that
\begin{equation}
\langle\exp(-\chi^\dagger H\otimes\openone_2\chi) \rangle_H =
\exp\left[-\frac{\lambda^2}{2N}\mbox{tr}(u^2)\right] \;,
\label{eq:apc4}
\end{equation}
where $u$ is the following $2\times 2$ matrix
\begin{equation}
u = \left(\begin{array}{cc} \chi_1^\dagger\chi_1 &
\chi_2^\dagger\chi_1 \\ \chi_1^\dagger\chi_2 & \chi_2^\dagger\chi_2
\end{array}\right).
\label{eq:apc5}
\end{equation}
The quartic term can be decoupled by a Hubbard-Stratonovich
transformation, namely,
\begin{equation}
\int d[Q] \exp\left[ -\frac{N}{2}\mbox{tr}(Q^2)- 
i\lambda\mbox{tr}(Qu)\right] = \left(\frac{2\pi}{N}\right)^2 
\exp\left[-\frac{\lambda^2}{2N}\mbox{tr}(u^2) \right],
\label{eq:apc6}
\end{equation}
where $Q$ is a $2\times 2$ Hermitian matrix and
$d[Q]=dQ_{11}dQ_{22}dQ_{12}dQ_{21}$. As a result, we have
\begin{equation}
f(a) = \left(\frac{N}{2\pi}\right)^2 \!\int \!d[Q] \exp\left[
    -\frac{N}{2}\mbox{tr}(Q^2)\right] \int\! d[\chi] \exp[
    -\chi^\dagger(\alpha \openone_N\otimes L+i\lambda
    \openone_N\otimes Q)\chi].
\label{eq:apc7}
\end{equation}
The Gaussian integral over the fermionic variables can be easily
carried out, yielding
\begin{equation}
f(a) = \left(\frac{N}{2\pi}\right)^2 \int d[Q] \exp\left\{
-\frac{N}{2}\mbox{tr}(Q^2) + N\mbox{tr}[\ln(\alpha L+i\lambda
Q)]\right\}.
\label{eq:apc8}
\end{equation}

When $N\gg 1$ the above integral over $Q$ can be evaluated by the
saddle-point approximation (which becomes exact in the limit
$N\rightarrow\infty$). For this purpose we first separate angular and
radial components of $Q$, namely, $Q=T^\dagger qT$, where $q=\mbox{
diag}(q_1,q_2)$ and $T$ is an SU(2) matrix. The differential breaks up
into $d[Q]=d\mu(T)J(q)d[q]$, where $J(q)=\pi(q_1-q_2)^2$ is the
Jacobian of the transformation, $d[q]=dq_1dq_2$, and $d\mu(T)$ is the
group measure normalized to unity. This yields
\begin{eqnarray}
f(a) & = & \left(\frac{N}{2\pi}\right)^2\int d[q] J(q) \exp(-NF[q])
\int d\mu(T) \exp\left[-\frac{i\alpha
N}{\lambda}\mbox{tr}(q^{-1}TLT^\dagger)\right],
\label{eq:apc9}
\end{eqnarray}
where $F[q] = (1/2)\mbox{tr}(q^2)-\mbox{tr}[\ln(i\lambda q)]$ and we
have only kept terms to lowest order in $\alpha$. The saddle-point
expansion now involves only the radial part of the action:
\begin{eqnarray}
F[q] & = & 1 - 2\ln(\lambda) - \frac{i\pi}{2}(\sigma_1+\sigma_2) +
\delta q_1^2 + \delta q_2^2 + O(\delta q^3),
\label{eq:apc10}
\end{eqnarray}
where $q_{1,2}=\sigma_{1,2}+\delta q_{1,2}$ and
$\sigma_{1,2}^2=1$. The relevant saddle points correspond to
$\sigma_1=-\sigma_2$, resulting in
\begin{eqnarray}
f(a) & = & N \left( \frac{\lambda^2}{e} \right)^{N}\! \int \!d\mu(T)
\exp \left[ -\frac{i\alpha\sigma_1N}{\lambda} \mbox{tr}(LTLT^\dagger)
\right].
\label{eq:apc11}
\end{eqnarray}
The integral over the SU(2) manifold can be evaluated through the
well-known Itzykson-Zuber formula,\cite{Itzykson80} which in a
simplified form reads
\begin{equation}
\int d\mu(T) \exp[\beta\mbox{tr}(LTLT^\dagger)] =
\frac{\det[\exp(\beta l_{i}l_{j})]}{\beta(l_1-l_2)^2},
\label{eq:apc12}
\end{equation}
with $l_{1,2}$ denoting the eigenvalues of $L$. Hence,
\begin{equation}
f(a) = 2N \left(\frac{\lambda^2}{e}\right)^{N} \frac{\sin(2\alpha
N/\lambda)}{(2\alpha N/\lambda)}
\label{eq:apc13}
\end{equation}
(the factor of 2 takes into account the double saddle point). Finally,
we obtain
\begin{equation}
c_N = \frac{4N^3}{3\lambda^2}\left(\frac{\lambda^2}{e}\right)^{N}\;.
\label{eq:apc14}
\end{equation}

\section{The asymptotic limits of $c_\Gamma(x)$}
\label{sec:appendixB}

The small-$X$ asymptotics of the correlator of decay widths can be
determined by the same method used in Sec.~\ref{sec:subcondpeakcor}
for the conductance peak height correlator. Beginning with the
definition presented in Eq.~(\ref{eq:corWW}), we expand
$\Gamma_{k\nu}(\bar{X}\pm X/2)$ up to first order in $X$ [see
Eq.~(\ref{eq:expgamma})]. The zeroth-order term of $C_\Gamma(X)$ is
then given by
\begin{equation}
C_\Gamma(0) = \langle \Gamma_\nu^2 \rangle - \langle \Gamma_\nu
\rangle^2 = \langle \Gamma \rangle^2
\label{eq:C0Gz}
\end{equation}
for the unitary ensemble. An expression analogous to
Eq.~(\ref{eq:expCX}) is used to write the second-order coefficient of
$C_\Gamma(X)$ in terms of the amplitudes $\Lambda_\nu$, namely,
\begin{equation}
C_\Gamma^{\prime\prime}(0) = -\frac{1}{2} \langle \Lambda_\nu^2
\rangle.
\label{eq:C0Gpp1}
\end{equation}
Carrying out the average over the matrix elements of the external
perturbation $U$ [see Eq.~(\ref{eq:averLambda1})], we find that
\begin{equation}
C_\Gamma^{\prime\prime}(0) = - \frac{\sigma^2}{N} \left\langle
\sum_{\mu\ne\nu} \frac{\Gamma_\nu\Gamma_\mu}{\varepsilon_{\nu\mu}^2}
\right\rangle.
\label{eq:C0Gpp2}
\end{equation}
We now average separately over the eigenvalues and partial widths and
obtain
\begin{equation}
C_\Gamma^{\prime\prime}(0) = - \frac{ 2\pi^2 \langle\Gamma\rangle^2
\sigma^2 X^2} {3N\Delta^2}.
\label{eq:C0Gpp3}
\end{equation}
Upon rescaling both $C_\Gamma(X) \rightarrow c_\Gamma(x) =
C_\Gamma(X)/\langle\Gamma\rangle^2$ and $X \rightarrow x =
X(\sigma^2/N\Delta^2)$, we arrive at
\begin{equation}
c_\Gamma(x) = 1 - \frac{2\pi^2 x^2}{3} + O(x^4).
\label{eq:C0Gpp4}
\end{equation}

The large-$X$ asymptotics of $C_\Gamma(X)$ can be inferred from the
asymptotics of another correlator, namely,
\begin{equation}
P(X,E) = \Omega^2 \left\langle \sum_{\mu,\nu} |\psi_\mu(r;X_1)|^2
|\psi_\nu(r;X_2)|^2 \delta\Bigl(E_1-\varepsilon_\mu(X_1)\Bigr)
\delta\Bigl(E_2-\varepsilon_\nu(X_2)\Bigr) \right\rangle -
\frac{1}{\Delta^2}
\label{eq:CGtilde} 
\end{equation}
where $\Omega$ is the system volume, $X_{1,2}=\bar{X}\mp X/2$, and
$E_{1,2}=\bar{E}\mp E/2$. Recall that the wave function intensities
are proportional to the decay widths $\Gamma_\mu(X_1)$ and
$\Gamma_\nu(X_2)$ for pointlike contacts. At $E_1=E_2$ and large $X$,
the interlevel correlations are secondary to intralevel ones; as a
result, the $\delta$ function in Eq.~(\ref{eq:CGtilde}) acts as a
Kronecker $\delta$, causing $P(X,0)$ and $C_\Gamma(X)$ to coincide (up
to a prefactor equal to $\Delta^2$) to leading order in $O(1/X)$.

Let us for convenience assume a finite size space basis to represent
the system Hamiltonian. We can then reduce Eq.~(\ref{eq:CGtilde}) to
\begin{equation}
P(X,E) = \left(\frac{N}{\pi}\right)^2 \left\langle
\mbox{Im}\left[G(E_1+i\epsilon;X_1)\right]_{kk}
\mbox{Im}\left[G(E_2-i\epsilon;X_2)\right]_{kk} \right\rangle -
\frac{1}{\Delta^2}
\label{eq:apd2}
\end{equation}
with $G(E;X)=[E-H(X)]^{-1}$ and $\epsilon\rightarrow 0^+$. The above
expression can be rewritten in the more convenient form
\begin{equation}
P(X,E) = -\left(\frac{N^2}{2\pi^2}\right) \mbox{Re} \left\langle
\left[G(E_1+i\epsilon;X_1)\right]_{kk}
\left[G(E_2-i\epsilon;X_2)\right]_{kk} \right\rangle -
\frac{1}{\Delta^2} \;.
\label{eq:apd3}
\end{equation}
In general, an expression like Eq.~(\ref{eq:apd3}) requires the
evaluation of the following quantity:
\begin{equation}
D_{klmn}(E,X) = \left\langle [G(E_1+i\epsilon;X_1)]_{kl}
[G(E_2-i\epsilon;X_2)]_{mn} \right\rangle.
\label{eq:apd4}
\end{equation}
The correlator $D_{klmn}(E,X)$ can be calculated exactly in the
zero-mode approximation of the supersymmetric nonlinear $\sigma$ model
\cite{Efetov83} (or, equivalently, in the RMT framework). This
calculation is standard nowadays (for a recent review, see
Ref.~\onlinecite{Fyodorov95}) and has already been presented in the
literature.\cite{Taniguchi93} Here we will only mention the resulting
expression for the unitary ensemble, which is
\begin{eqnarray}
D_{klmn}(E,X) & = & \left(\frac{\pi}{N\Delta}\right)^2 \left[
\delta_{kl}\delta_{mn} - \delta_{kl}\delta_{mn} k(\omega,x)
\right. \nonumber \\ & & \left. - \delta_{kn}\delta_{lm} n(\omega,x)
\right] \;,
\label{eq:apd21}
\end{eqnarray}
where
\begin{eqnarray}
k(\omega,x) & = & \int_1^\infty d\lambda_1 \int_{-1}^1 d\lambda_2
\exp\left[ 2\pi i(\omega/2+i\eta)(\lambda_1-\lambda_2) -
(\pi^2x^2/2)(\lambda_1^2-\lambda_2^2) \right]
\label{eq:apd22}
\end{eqnarray}
and
\begin{equation}
n(\omega,x) = \int_1^\infty d\lambda_1 \int_{-1}^1 d\lambda_2
\left(\frac{\lambda_1+\lambda_2}{\lambda_1-\lambda_2}\right)
\exp\left[ 2\pi i(\omega/2+i\eta)(\lambda_1-\lambda_2) -
(\pi^2x^2/2)(\lambda_1^2-\lambda_2^2) \right].
\label{eq:apd23}
\end{equation}
When writing these equations we have rescaled the variables to
$E/\Delta=\omega$, $\epsilon/\Delta=\eta$, and
$N\Delta\sqrt{\mbox{tr}(U^2)}X/\pi^2=x$. Going back to
Eq.~(\ref{eq:apd3}), we arrive at
\begin{equation}
p(x,\omega) = \Delta^2 P(E,X) = \frac{1}{2}\mbox{Re}\left[ k(\omega,x)
+ n(\omega,x) \right] .
\label{eq:apd24}
\end{equation}
Since $k(0,x) \rightarrow 2/(\pi x)^4$ and $n(0,x)\rightarrow 2/(\pi
x)^2$ as $x\rightarrow\infty$, we have that $p(x,0) \rightarrow 1/(\pi
x)^2$ in the same limit. Therefore, we expect that
\begin{equation}
c_\Gamma(x) \stackrel{x\rightarrow\infty}{\longrightarrow}
\frac{1}{(\pi x)^2}.
\label{eq:apd27}
\end{equation}
Finally, we remark that the $x\gg1$ universal asymptotics of both
$k(0,x)$ and $n(0,x)$ can also be obtained by the diagrammatic
perturbation theory of disordered metals expressed in terms of
diffuson modes.

\section{Evaluation of $B(\lowercase{s})$}
\label{sec:appendixC}

In this Appendix we give a schematic description of the calculation of
$B(s)$. The starting point is Eq.~(\ref{eq:eigaver}). Here we go
through the same steps of Sec.~\ref{sec:subcondpeakcor} to evaluate
$A_\varepsilon$ [see Eq.~(\ref{eq:Beig})]. First we fix the reference
eigenvalue $\varepsilon_\nu$ to the center of the spectrum, obtaining
\begin{equation}
{\widetilde B}(k) = \frac{1}{\rho(0)} \left\langle \sum_{\nu=1}^N
\prod_{\mu\ne\nu} \delta(\varepsilon_\nu) \left[ 1 +
\frac{(k\Delta/2\pi)^2} {\varepsilon_{\nu\mu}^2} \right]^{-1}
\right\rangle_{\{\varepsilon_\nu\}}\;.
\label{eq:appc1}
\end{equation}
Next, we rephrase this expression in terms of an average over a
spectral determinant, namely,
\begin{equation}
{\widetilde B}(k) = a_N \left\langle \frac{\det \left( H^4 \right)}
{\det \left[ H^2 + (k\Delta/2\pi)^2 \right]} \right\rangle_H,
\label{eq:appc2}
\end{equation}
where $a_N$ is a constant [such that ${\widetilde B}(0)=1$] and the
average is performed over a $(N-1)\times(N-1)$ GUE matrix $H$. The
appearance of determinants in both numerator and denominator in
Eq.~(\ref{eq:appc2}) makes its evaluation technically more difficult
than Eq.~(\ref{eq:apc1}). It is necessary to introduce not only four
anticommuting auxiliary variables, but also two commuting (complex)
ones. The resulting symmetry group is U$(1,1|4)$ (the pseudo-unitarity
is due to the structure of the denominator). Fortunately, a general
solution for such graded symmetry problems has been recently worked
out.\cite{Andreev95} The derivation is a nontrivial generalization of
the method of Appendix \ref{sec:appendixA}. For an expression with the
structure of Eq.~(\ref{eq:appc2}), one arrives at the following
formula:\cite{Andreev95}
\begin{eqnarray}
\left\langle \frac{\prod_{j=1}^4 \det(H - m_j\Delta)} {\det(H -
i\alpha\Delta) \det(H + i\alpha\Delta)} \right\rangle_H & = & \frac{A
e^{-2\pi\alpha}}{\alpha} \sum_{\{m_j\}} \Biggl[ \frac{(i\alpha-m_3)
(i\alpha-m_4)} {(m_3-m_1) (m_3-m_2)} \nonumber \\ & & \times
\frac{(i\alpha+m_1) (i\alpha+m_2)} {(m_4-m_1) (m_4-m_2)}
e^{i\pi(m_1+m_2-m_3-m_4)} \Biggr],
\label{Eq:apcc3}
\end{eqnarray}
where the sum runs over all six nonequivalent combinations of pairs of
$m_j$, $\alpha>0$, and $A$ is an unspecified constant. To get
${\widetilde B}(k)$ we need to take the limit $m_j\rightarrow 0$ for
all $j=1,2,3,4$ at a given order. After some algebra, one finds that
\begin{equation}
{\widetilde B}(k) = \frac{e^{-k}}{24} \left( 24 + 24k +8k^2 +k^3
\right).
\label{eq:apcc4}
\end{equation} 
Finally, inverse Fourier transforming the above expression, we arrive
at
\begin{eqnarray}
B(s) & = & \int_{-\infty}^{\infty} \frac{dk}{2\pi}\ e^{-iks}
{\widetilde B}(k) \nonumber \\ & = & \frac{35 + 14s^2 +
3s^4}{12\pi(1+s^2)^4}.
\label{eq:apcc5}
\end{eqnarray}



\newpage

\newpage

\newpage

\newpage

\newpage

\newpage

\newpage

\newpage

\newpage

\begin{thebibliography}{99}


\bibitem{ChaosReview} Reviews of experimental and theoretical aspects
of quantum chaos in nanostructures can be found in Chaos {\bf 3},
(1993). See also H. U. Baranger, in {\it Nanotechnology}, edited by
G.\ Timp (AIP, New York, in press); R. M. Westervelt, {\it ibid.}

\bibitem{Jalabert90} R. A. Jalabert, H. U. Baranger, and A. D. Stone,
Phys. Rev. Lett.  {\bf 65}, 2442 (1990).
    
\bibitem{Marcus92} C. M. Marcus, A. J. Rimberg, R. M. Westervelt,
P. F. Hopkins, and A. C. Gossard, Phys. Rev. Lett. {\bf 69}, 506
(1992).

\bibitem{Kel94}
M. W. Keller, O. Millo, A. Mittal, D. E. Prober, and R. N. Sacks,
Surf. Sci. {\bf 305}, 501 (1994).

\bibitem{Chang94}
A. M. Chang, H. U. Baranger, L. N. Pfeiffer, and K. W. West,
Phys. Rev. Lett.  {\bf 73}, 2111 (1994).

\bibitem{MJBerry94}
M. J. Berry, J. A. Katine, R. M. Westervelt, and A. C. Gossard,
Phys. Rev. B {\bf 50}, 17 721 (1994).

\bibitem{Bird94}
J. P. Bird, K. Ishibashi, Y. Aoyagi, T. Sugano, and Y. Ochiai,
Phys. Rev. B {\bf 50}, 18 678 (1994).

\bibitem{Kastner92} For a review, see M. A. Kastner,
Rev. Mod. Phys. {\bf 64}, 849 (1992), and references therein.

\bibitem{Sivan94} 
See, for instance, U. Sivan, F. P. Milliken, K. Milkove, S. Rishton,
Y. Lee, J. M. Hong, V. Boegli, D. Kern, and M. deFranza,
Europhys. Lett. {\bf 25}, 605 (1994).

\bibitem{Jalabert92} 
R. A. Jalabert, A. D. Stone, and Y. Alhassid, Phys. Rev. Lett. {\bf
68}, 3468 (1992).

\bibitem{Prigodin93}
V. N. Prigodin, K. B. Efetov, and S. Iida, Phys. Rev. Lett. {\bf 71},
1230 (1993).

\bibitem{Bruus94} H. Bruus and A. D. Stone, Phys. Rev. B {\bf 50}, 18
275 (1994); Physica B {\bf 189}, 43 (1993).

\bibitem{Mucciolo95a}
E. R. Mucciolo, V. N. Prigodin, and B. L. Altshuler, Phys. Rev. B {\bf
51}, 1714 (1995).

\bibitem{Alhassid95} Y. Alhassid and C. H. Lewenkopf,
Phys. Rev. Lett. {\bf 75}, 3922 (1995).

\bibitem{Baranger94} 
H. U. Baranger and P. A. Mello, Phys. Rev. Lett. {\bf 74}, 142 (1994);
R. A. Jalabert, J.-L. Pichard, and C. W. J. Beenakker,
Europhys. Lett. {\bf 27}, 255 (1994).

\bibitem{Chang95}
A. M. Chang, H. U. Baranger, L. N. Pfeiffer, K. W. West, and
T. Y. Chang, Phys. Rev. Lett. {\bf 76}, 1695 (1996).

\bibitem{Brouwer95}
P. M. Brouwer and C. W. J. Beenakker, Phys. Rev. B {\bf 51} 7739
(1995).

\bibitem{Klein95} 
O. Klein (private communication).

\bibitem{Folk95} J.A. Folk, S. R. Patel, S. F. Godijn, A. G. Huibers,
S. M. Cronenwett, C. M. Marcus, K. Campman, and A. C. Gossard,
Phys. Rev. Lett {\bf 76}, 1699 (1996).

\bibitem{Chan95}
I. H. Chan, R. M. Clarke, C. M. Marcus, K. Campman, and A. C. Gossard,
Phys. Rev. Lett. {\bf 74}, 3876 (1995).

\bibitem{Bohigas91}
O. Bohigas, in {\it Chaos and Quantum Physics}, edited by M.-J.
Giannoni, A. Voros, and J. Zinn-Justin (North-Holland, Amsterdam,
1991).

\bibitem{Obs1} 
For classical hyperbolic systems one expects quantum universal
behavior in the semiclassical region of the spectrum, i.e., where the
electron wavelength at the Fermi surface is small enough to resolve
the typical ``irregularities" of the cavity. This is usually the case
for two-dimensional systems with more than $O(10^2)$ electrons.

\bibitem{Landauer57}
R. Landauer, IBM J. Res. Dev. {\bf 1}, 223 (1957).

\bibitem{Buettiker}
M. B\"uttiker, Phys. Rev. Lett. {\bf 57}, 1761 (1986); IBM
J. Res. Dev. {\bf 32}, 317 (1988).

\bibitem{Lewenkopf91}
C. H. Lewenkopf and H. A. Weidenm\"uller, Ann. Phys. (N.Y.) {\bf 212},
53 (1991).

\bibitem{Lane58}
A. M. Lane and R. G. Thomas, Rev. Mod. Phys. {\bf 30}, 257 (1958).

\bibitem{Zirnbauer93}
M. R. Zirnbauer, Nucl. Phys. {\bf A560}, 95 (1993).

\bibitem{Meir91}
Y. Meir, N. S. Wingreen, and P. A. Lee, Phys. Rev. Lett. {\bf 66},
3048 (1991).

\bibitem{Beenakker91}
C. W. J. Beenakker, Phys. Rev. B {\bf 44}, 1646 (1991).

\bibitem{Kleintip}
We thank O. Klein for pointing this out to us.

\bibitem{Mehta91}
M. L. Mehta, {\it Random Matrices}, 2nd ed. (Academic Press, San
Diego, 1991).

\bibitem{Szafer93}
A. Szafer and B. L. Altshuler, Phys. Rev. Lett. {\bf 70}, 587 (1993).

\bibitem{Simons93a}
B. D. Simons and B. L. Altshuler, Phys. Rev. B {\bf 48}, 5422 (1993).

\bibitem{Uaverage} 
The averaging is done explicitly here to simplify the calculation. In
principle $U$ is not required to be a random matrix. However, once we
choose for basis the eigenstates of $H_0$, the matrix elements
$U_{\nu\mu}$ will be Gaussian distributed, regardless of the nature of
$U$.

\bibitem{Porter65} C. E. Porter, in {\it Statistical Theories of
Spectra: Fluctuations}, edited by C. E. Porter (Academic Press, New
York, 1956).

\bibitem{Itzykson89} C. Itzykson and J.-M. Drouffe, {\it Statistical
Field Theory} (Cambridge University Press, Cambridge, England, 1989),
Vol. 2.

\bibitem{Alhassid94}
A similar correlator was previously considered by Y. Alhassid and
H. Attias, Phys. Rev. Lett {\bf 74}, 4365 (1994).

\bibitem{GOE} 
By extending the arguments of Appendix B to the GOE, one finds that
$c_\Gamma(x)\rightarrow 2/(\pi x)^2$ when $x\rightarrow\infty$.

\bibitem{Mucciolo95b}
E. R. Mucciolo, B. D. Simons, A. V. Andreev, and V. N. Prigodin,
Phys. Rev. Lett. {\bf 75}, 1360 (1995).

\bibitem{Andreev95}
A. V. Andreev and B. D. Simons, Phys. Rev. Lett. {\bf 75}, 2304
(1995).

\bibitem{Berry86}
M. V. Berry and M. Robnik, J. Phys. A {\bf 19}, 649 (1986).

\bibitem{Berry94}
M. V. Berry and J. P. Keating, J. Phys. A {\bf 27}, 6167 (1994).

\bibitem{Almeida91}
A. M. Oz\'orio de Almeida, in {\it Quantum Chaos}, edited by
H. A. Cerdeira, R. Ramaswamy, M. C. Gutzwiller, and G. Casati (World
Scientific, Singapore, 1991).

\bibitem{Bohigas95}
O. Bohigas, M.-J. Giannoni, A. M. Oz\'orio de Almeida, and C. Schmit,
Nonlinearity {\bf 8}, 203 (1995).

\bibitem{Hannay84}
J. H. Hannay and A. M. Oz\'orio de Almeida, J. Phys. A {\bf 17}, 3429
(1984).

\bibitem{Pluhar95}
Z. Pluha\v{r}, H. A. Weidenm\"uller, J. A. Zuk, C. H. Lewenkopf, and
F. J. Wegner, Ann. Phys. (N.Y.) {\bf 243}, 1 (1995).

\bibitem{Obs3} The decay of classical autocorrelators for ergodic
dynamical systems is presently a matter of intensive investigation. In
the absence of a rigorous proof we base our statement about the
convergence of $\int_0^T\!dt \, C(t)$ on numerical evidence.

\bibitem{Obs4} The underlying idea is again the uniform coverage of
the phase space. For an ergodic system the typical winding number as a
function of length should not depend on whether the trajectory is
periodic or not, provided that $L>L_c$.

\bibitem{Obs5} 
An example of this procedure for the case of ballistic quantum dots in
the presence of high magnetic fields is given by P. L. McEuen {\it et
al.}, Phys. Rev. B {\bf 45}, 11 419 (1992). For heavily doped,
diffusive dots, a single-particle description of the spectrum is
probably insufficient (see Ref.\onlinecite{Sivan94}).

\bibitem{Itzykson80}
C. Itzykson and J. B. Zuber, J. Math. Phys. {\bf 21}, 411 (1980).

\bibitem{Efetov83}
K. B. Efetov, Adv. Phys. {\bf 32}, 53 (1983).

\bibitem{Fyodorov95} 
Y. V. Fyodorov, in {\it Mesoscopic Quantum Physics}, edited by
E. Akkermans, G. Montambaux, J.-L. Pichard, and J. Zinn-Justin
(North-Holland, Amsterdam, in press).

\bibitem{Taniguchi93} 
N. Taniguchi, A. V. Andreev, and B. L. Altshuler, Europhys.
Lett. {\bf 29}, 515 (1995).

\bibitem{Alhassid96}
Y. Alhassid and H. Attias, Phys. Rev. Lett. {\bf 76}, 1711 (1996).

\end{thebibliography}
\end{document}